\documentclass{aa}

\usepackage[varg]{txfonts}
\usepackage[export]{adjustbox}
\usepackage{natbib}
\usepackage{blindtext}
\usepackage{booktabs}
\usepackage{pifont}
\usepackage{lscape}
\usepackage{multirow}
\usepackage{tabularx}
\usepackage{pbox}
\usepackage{wasysym}
\usepackage{amsmath}
\usepackage[capitalise]{cleveref}
\usepackage{upgreek}
\usepackage{listings}
\usepackage{graphics}
\usepackage{color}

\begin{document}

\title{VLBA polarimetric monitoring of 3C 111}

\author{T.~Beuchert\inst{\ref{inst1},\ref{inst2},\thanks{Now at Anton
  Pannekoek Institute for Astronomy,  P.O. Box 94249, 1090GE Amsterdam; contact: t.beuchert@uva.nl}}
\and M.~Kadler\inst{\ref{inst2}}
\and M.~Perucho\inst{\ref{inst3},\ref{inst4}}
\and C.~Gro{\ss}berger\inst{\ref{inst5}}
\and R.~Schulz\inst{\ref{inst6},\ref{inst2}}
\and I.~Agudo\inst{\ref{inst7}}
\and C.~Casadio\inst{\ref{inst8},\ref{inst7}}
\and J.~L.~G\'{o}mez\inst{\ref{inst7}} 
\and M.~Gurwell\inst{\ref{inst9}}
\and D.~Homan\inst{\ref{inst10}}
\and Y.~Y.~Kovalev\inst{\ref{inst11},\ref{inst12},\ref{inst8}}
\and M.~L.~Lister\inst{\ref{inst13}}
\and S.~Markoff\inst{\ref{inst14}}
\and S.~N.~Molina\inst{\ref{inst7}}
\and A.~B.~Pushkarev\inst{\ref{inst15},\ref{inst11}}
\and E.~Ros\inst{\ref{inst8},\ref{inst3},\ref{inst4}}
\and T.~Savolainen\inst{\ref{inst16},\ref{inst17},\ref{inst8}}
\and T.~Steinbring\inst{\ref{inst2}}
\and C.~Thum\inst{\ref{inst18}}
\and J.~Wilms\inst{\ref{inst1}}
}

\institute{ Dr.\ Remeis-Sternwarte \& Erlangen Centre for
  Astroparticle Physics, Universit\"at Erlangen-N\"urnberg,
  Sternwartstrasse 7, 96049 Bamberg, Germany
  \label{inst1}
  \and Lehrstuhl f\"ur Astronomie, Universit\"at W\"urzburg,
  Emil-Fischer-Str. 31, 97074 W\"urzburg, Germany
  \label{inst2}
  \and Departament d'Astronomia i Astrof\'isica, Universitat de
  Val\`encia C/ Dr. Moliner 50, 46100 Burjassot, Val\`encia, Spain
  \label{inst3}
  \and Observatori Astron\`omic, Universitat de Val\`encia, Parc
  Científic, C. Catedr\'atico Jos\'e Beltr\'an 2, 46980 Paterna,
  Val\`encia, Spain
  \label{inst4}
  \and Max-Planck-Institut für extraterrestrische Physik (MPE), PO
  1312, 85741 Garching, Germany
  \label{inst5}
  \and Netherlands Institute for Radio Astronomy (ASTRON), PO Box 2,
  7990 AA Dwingeloo, The Netherlands
  \label{inst6}
  \and Instituto de Astrof\'isica de Andaluc\'ia-CSIC, Apartado 3004,
  18080, Granada, Spain
  \label{inst7}
  \and Max-Planck-Institut für Radioastronomie, Auf dem Hügel 69,
  53121, Bonn, Germany
  \label{inst8}
  \and Harvard-Smithsonian Center for Astrophysics, 60 Garden St.,
  Cambridge, MA 02138, USA
  \label{inst9}
  \and Department of Physics, Denison University, 100 W College St.,
  Granville, OH 43023, USA
  \label{inst10}
  \and Astro Space Center of Lebedev Physical Institute, Profsoyuznaya
  84/32, 117997 Moscow, Russia 
  \label{inst11}
  \and Moscow Institute of Physics and Technology, Dolgoprudny,
  Institutsky per., 9, 141700 Moscow region, Russia
  \label{inst12}
  \and Department of Physics and Astronomy, Purdue University, 525
  Northwestern Avenue, West Lafayette, IN 47907, USA
  \label{inst13}
  \and Anton
  Pannekoek Institute for Astronomy, P.O. Box 94249, 1090GE
  Amsterdam, The Netherlands
  \label{inst14}
  \and Crimean Astrophysical Observatory, 98409 Nauchny, Crimea, Russia
  \label{inst15}
  \and Aalto University Department of Electronics and Nanoengineering,
  PL 15500, 00076, Aalto, Finland
  \label{inst16}
  \and Aalto University Mets\"ahovi Radio
  Observatory, Mets\"ahovintie 114, 02540 Kylm\"al\"a, Finland
  \label{inst17}
  \and Instituto de Radio
  Astronom\'ia Milim\'etrica, Avenida Divina Pastora, 7, Local 20,
  18012 Granada, Spain
  \label{inst18}
}

\date{14-09-2017 /
30-10-2017 }

\abstract {While studies of large samples of jets of Active Galactic
  Nuclei (AGN) are important in order to establish a global picture,
  dedicated single-source studies are an invaluable tool for probing
  crucial processes within jets on parsec scales. These processes
  involve in particular the formation and geometry of the jet magnetic
  field as well as the flow itself.}{We aim to better understand the
  dynamics within relativistic magneto-hydrodynamical flows in the
  extreme environment and close vicinity of supermassive black
  holes.}{In order to follow that aim, we analyze the peculiar radio
  galaxy 3C~111, for which long-term polarimetric observations are
  available. We make use of the high spatial resolution of the VLBA
  network and the MOJAVE monitoring program, which provides
  high data quality also for single sources
  and allows us to study jet dynamics on parsec scales in full
  polarization with an evenly sampled time-domain. While electric
  vectors can probe the underlying magnetic field, other properties of
  the jet such as the variable (polarized) flux density, feature size
  and brightness temperature, can give valuable insights into the flow
  itself. We complement the VLBA data with data from the IRAM 30-m
  Telescope as well as the SMA.}{We observe a complex evolution of the
  polarized jet. The electric vector position angles (EVPAs) of
  features traveling down the jet perform a large rotation of
  $\gtrsim$180\degr\ across a distance of about 20\,pc. As opposed to
  this smooth swing, the EVPAs are strongly variable within the first
  parsecs of the jet. We find an overall tendency towards transverse
  EVPAs across the jet with a local anomaly of aligned vectors in
  between. The polarized flux density increases rapidly at that
  distance and eventually saturates towards the outermost observable
  regions. The transverse extent of the flow suddenly decreases
  coincident with a jump in brightness temperature around where we
  observe the EVPAs to turn into alignment with the jet flow. Also the
  gradients of the feature size with distance and the particle density
  steepen significantly at that region.}{We interpret the propagating
  polarized features with shocks and the observed local anomalies with
  the interaction of these shocks with a localized recollimation shock
  of the underlying flow. Together with a sheared magnetic field, this
  shock-shock interaction can explain the large rotation of the
  EVPA. The superimposed variability of the EVPAs close to the core is
  likely related to a clumpy Faraday screen, which also contributes
  significantly to the observed EVPA rotation in that region.}


\maketitle

\section{Introduction}\label{sec:intro}
Collimated jet outflows have been observed in many Active Galactic
Nuclei (AGN). They may either be launched by magneto-centrifugal
forces \citep{BP1982}, by the extraction of energy from a Kerr black
hole \citep{BZ1977} or a combination of both. Both predict increasing
magnetization of the produced jet towards the black hole. Modern GRMHD
simulations extend on these formalisms and turn out to be good
descriptions of observed jets
\citep[e.g.,][]{McKinney2004,Tchekhovskoy2011}.

Such jets have been shown to be strong emitters across the
electromagnetic spectrum \citep[e.g.,][for the kpc-scale jet of
3C~111]{Clautice2016}. In the radio band, the technique of VLBI can
resolve jets on parsec scales. In blazars, distinct features are
typically found to be ejected at apparent superluminal velocities
\citep[e.g.,][]{Lister2013} and emit a non-thermal synchrotron
spectrum, reflecting the presence of relativistic charged particles in
a magnetic field that is at least partially ordered at the location of
these features.

Polarization sensitive observations of parsec-scale jets have
confirmed this notion and found many of these features to be
significantly polarized \citep[e.g.,][]{Homan2005}. The degree of
polarization typically lies around a few percent in the
subparsec-scale core region and increases to higher levels in the jet
further downstream, often coinciding with propagating features that
are commonly interpreted as relativistic shocks
\citep{Koenigl1981,Marscher1985}, which can enhance the total and
polarized flux density.  Polarimetry is therefore a valuable tool to
also study the magnetic field structure inside jets. Studies of large
samples of blazars seem to confirm a quasi-bimodal distribution of
EVPAs with the majority of VLBI knots in BL~Lac objects showing
aligned EVPAs, while this picture is not as clear for quasars
\citep{Gabuzda1994,Gabuzda2000,Lister2000,Lister2005,Pollack2003}. \citet{Wardle1998},
\citet{Wardle2013} and \citet{Homan2005} emphasize a non-negligible
fraction of oblique EVPA orientations, i.e., ``local anomalies'', in
quasar jets as opposed to clear bimodal distributions (Agudo et
al.~2017b, {MNRAS}, subm.). For the simplified case of an underlying
axisymmetric and helical magnetic field, these results would imply a
dominance of toroidally dominated fields in the inner jet of BL~Lac
objects \citep{Lyutikov2005}, and a range of field directions in
quasars, e.g., due to oblique shocks \citep{Marscher2002}. At larger
distances, this picture may change: FR~II radio galaxies, the likely
parent populations of quasars, in general tend to produce
perpendicular EVPAs farther downstream
\citep{Bridle1984a,Cawthorne1993a,Bridle1994}, favoring the
interpretation with strong axial field components.

While surveys of the polarization properties are important for the
bigger picture, dedicated studies of the dynamics of spatially
resolved polarized parsec-scale jets are invaluable to gain more
insight into the complex processes inherent to these systems.
Steady-state (general relativistic) magnetohydrodynamical ((G,R)MHD)
simulations predict the formation of recollimation shocks in
overpressured jets
\citep[e.g.,][]{Gomez1997,Mimica2009,RocaSogorb2009,Fromm2016,Marti2016}. Observations
were able to confirm such recollimation shocks for BL~Lac
\citep{Cohen2014,Gomez2016}, CTA~102 \citep{Fromm2013,Fromm2016} and
3C~120 \citep{LeonTavares2010,Agudo2012}. Recollimation shocks have
also been associated with stationary features observed in radio
galaxies and blazars close to the radio core at millimeter wavelengths
\citep{Jorstad2001,Kellermann2004,Jorstad2005,Britzen2010,Jorstad2013}.

The evidence for recollimations close to the radio core fits into the
frame of an early paradigm of a ``master'' recollimation
\citep{Marscher2006,Marscher2008nat,Marscher2010} limiting the
acceleration and collimation zone
\citep{Beskin1998,Beskin2006,Meier2013}. This paradigm goes back to
work by \citet{Daly1988} and \citet{Impey1988}, who show that the core
at millimeter wavelengths may not only be the $\tau=1$ surface but a
characteristic feature, coinciding with a standing shock.
\begin{figure*}
\sidecaption
\includegraphics[width=12cm]{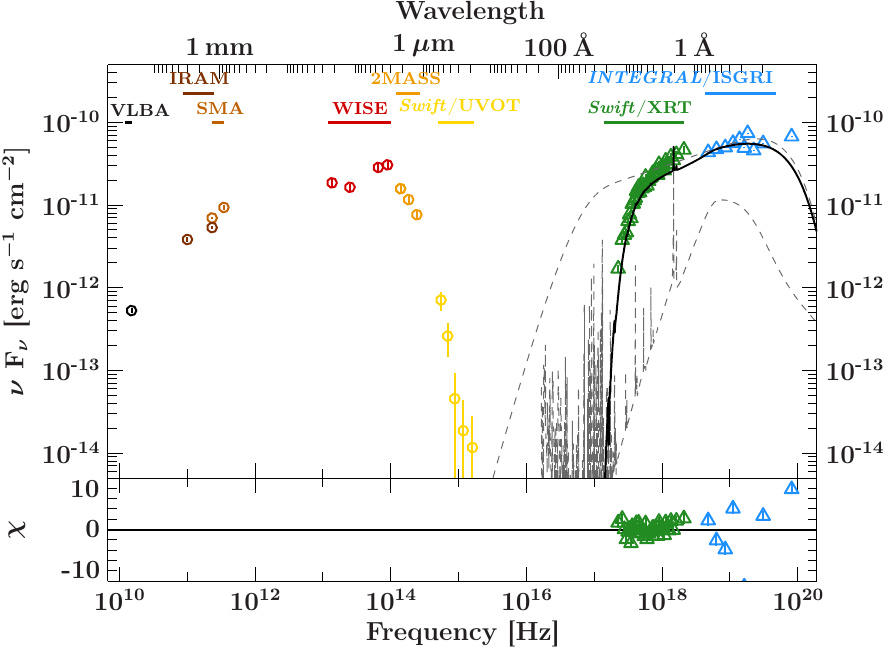}
\caption{Broad-band spectral energy distribution (SED) of 3C~111. The
  VLBA, IRAM and SMA flux-density points are weighted averages over
  the time range probed in this paper (2007--2012). We show integrated
  core+jet flux densities from the VLBA and the SMA.  The WISE
  \citep{Wright2010} and 2MASS \citep{Skrutskie2006} data are
  extracted from the All-Sky Source Catalog. The UV and X-ray data
  result from observations with \textit{Swift}/UVOT and XRT
  \citep{Burrows2005} on 2008-11-16. The hard X-ray
  \textit{INTEGRAL}/IBIS/ISGRI spectrum has been obtained in the
  time-interval considered in this paper using the HEAVENS online tool
  \citep{Walter2010}. The black, solid curve represents a model of an
  incident power-law (upper dashed curve) that is getting reprocessed
  in an accretion disk (lower dashed curve with prominent emission
  features). The residuals corresponding to the (hard) X-ray model are
  shown in the bottom panel. We can fit the XRT data well in
  combination with reflected emission from the accretion disk using
  \texttt{xillver} \citep{Garcia2013} and find
  $\chi\,\mathrm{(dof)}=47\,(34)$. The statistics worsen when
  including the \textit{INTEGRAL} data due to their scatter and
  systematic uncertainties that are not included here.}
\label{fig:sed}
\end{figure*}

\citet{Fromm2016} show for the case of CTA~102 that shocks propagating
downstream will eventually interact with such a standing, conical
recollimation shock. As \citet{Cawthorne2006} outlines, conical
recollimation shocks themselves can reveal a characteristic structure
of EVPAs, which matches observations of the core of the blazar
S5~1803$+$784 \citep{Cawthorne2013} as well as a downstream feature in
3C~120 \citep{Agudo2012}.  The polarization signatures of shock-shock
interactions are, however, unclear, and have not yet been consistently
quantified.

Meanwhile, systematic observations of blazars and radio galaxies
report smooth rotations of the EVPA with time (Agudo et al.~2017b,
{MNRAS}, subm.). Sample studies of blazars at optical wavelengths
\citep[e.g.,][]{Blinov2016} show that the amplitude of the rotations
is typically very large, reaching more than 180\degr\ on comparatively
short time scales. Prominent examples are BL~Lac
\citep{Marscher2008nat}, PKS~1510$-$089 \citep{Marscher2010} and
3C~279 \citep{Larionov2008,Abdo2010,Kiehlmann2016}. In the radio band,
however, large rotations of more than 90\degr are rare
\citep[e.g.,][]{Aller2003} and only found in a few sources
\citep[e.g.,][including BL~Lac and
PKS~1510$-$089]{Altschuler1980,Aller1981,Homan2002,Myserlis2016}. \citet{Marscher2010}
explain the large EVPA swing in PKS~1510$-$089 in terms of a
projected, geometrical rotation in presence of an ordered, helical
field observed at a shallow angle
\citep{Larionov2008,Nalewajko2010}. For the same source,
\citet{Myserlis2016} also find consistency with transitions between an
optically thin and thick state of the polarized ejecta. As
\citet{Homan2002} show, the EVPA swings of the majority of 12 selected
blazars are overall limited by 90\degr, consistent with changes of the
underlying magnetic field from toroidally dominated to poloidally
dominated or vice versa.  These observations demonstrate the ambiguity
when attempting to interpret polarimetric data.

Here, we are studying the long-term evolution of the
parsec-scale jet of the AGN of 3C~111 that shows particularly bright
and highly polarized traveling features. Our immediate aim is to study
the interaction of moving shocks with a stationary recollimation
shock. 3C~111 \citep[$z\sim 0.048$,][]{Veron2010} is classified as a
FR~II radio galaxy \citep{Sargent1977}.  Its extended twin
jet-structure \citep{Linfield1984,Leahy1997} is inclined with a PA of
about 63\degr.  Its radio core is exceptionally compact and
bright. The parsec-scale jet reveals features that undergo apparent
superluminal motion and appears as one-sided due to beamed emission
\citep{Jorstad2005,Kadler2008,Chatterjee2011}. These blazar-like
properties contrast the morphology on larger scales, which is
reminiscent of a typical radio galaxy. Long-term VLBI monitoring at
15\,GHz by
MOJAVE\footnote{\url{http://www.physics.purdue.edu/MOJAVE/}} reveals
strong structural variability on parsec scales as it has been observed
in total and polarized intensity by \citet{Kadler2008}, hereafter K08,
\citet{Grossberger2012} and \citet{Lister2013}. 3C~111 was also
subject to monitoring with the VLBA at 43\,GHz between 1998 and 2001
\citep{Jorstad2005} and between 2004 and 2010
\citep{Chatterjee2011}. All these studies revealed individual
ballistic components with apparent superluminal speeds of up to 6\,c,
most of which are significantly polarized.

At higher energies, however, multiwavelength studies
\citep{Chatterjee2011,Tombesi2012} reveal that the emission is more
reminiscent of a Seyfert galaxy, which makes 3C~111 a unique source to
study the disk-jet connection. Its broad-band spectral energy
distribution (SED) shows double-humped emission up to GeV energies
\citep[][and Fig.~\ref{fig:sed} for a version that is compiled with
radio, UV and X-ray data from within the time-interval considered in
this paper]{Hartman2008}.  While the soft energy hump is consistent
with synchrotron emission, the high-energy emission can be described
with a power law resulting from thermal and/or non-thermal
Comptonization in a compact corona \citep{deJong2012,Tombesi2013}. 
\begin{figure*}
\centering
\includegraphics[width=17cm]{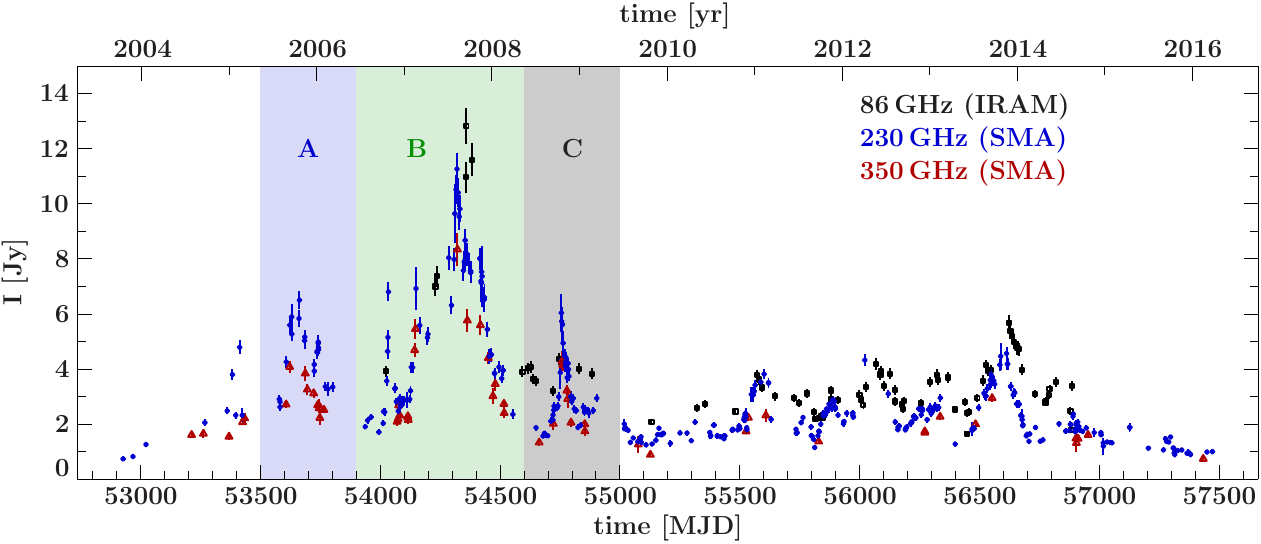}
\caption{Millimeter light curves of 3C~111 at 86\,GHz (IRAM, black
  squares), 230\,GHz, and 350\,GHz (SMA, blue diamonds and dark-red
  triangles, respectively). The shaded regions mark the three major
  outbursts A, B and C.}
\label{fig:mm_lc}
\end{figure*}

In the following, we are studying 36 MOJAVE epochs between 2007 and
2012 both in total and polarized intensity. This paper is structured
as follows. In Sect.~\ref{sec:obs} we outline the observational
techniques including single-dish instruments at millimeter wavelengths
and VLBI data at 15\,GHz. In Sect.~\ref{sec:results} the results are
described for the dynamics of the polarized parsec-scale jet, while
Sect.~\ref{sec:discussion} provides corresponding interpretations. A
summary is presented in Sect.~\ref{sec:summary}. Appendix A includes a
discussion on the viewing angle and describes a toy model facilitating
the interpretation on the complex processes taking place in the
polarized parsec-scale jet of 3C~111. Appendix B complements the
discussion with respect to the option of a helical field geometry
threading the jet of 3C~111.

The mass of the central black hole (BH) has
been derived to be $1.8_{-0.4}^{+0.5} \times 10^{8}\,M_{\astrosun}$
\citep{Chatterjee2011}. We use the latest cosmological parameters
provided by the \citet{Planck2016}, i.e., $\Omega_\mathrm{m}=0.308$,
$\Omega_{\uplambda}=0.692$, and
$H_{0}=67.8\,\mathrm{km}\,\mathrm{s}^{-1}\,\mathrm{Mpc}^{-1}$ and find
a correspondence of $1\,\mathrm{pc}/1\,\mathrm{mas}=1.08$.

\section{Observations and Data Analysis}
\label{sec:obs}

\subsection{SMA}
The 230\,GHz and 350\,GHz flux density data were obtained at the
Submillimeter Array (SMA) near the summit of Mauna Kea
(Hawaii). 3C~111 is included in an ongoing monitoring program at the
SMA to determine flux densities of compact extragalactic radio sources
that can be used as calibrators at millimeter wavelengths
\citep{Gurwell2007}. Observations of available potential calibrators
are from time to time observed for 3 to 5 minutes, and the measured
source signal strength calibrated against known standards, typically
solar system objects (Titan, Uranus, Neptune, or Callisto).  Data from
this program are updated regularly and are available at the SMA
website\footnote{\url{http://sma1.sma.hawaii.edu/callist/callist.html}}.

\subsection{IRAM 30-m}
We use data at 86\,GHz and 230\,GHz that were recorded with the
correlation polarimeter XPOL \citep{Thum2008} of the 30-m IRAM
Telescope on Pico Veleta (Spain). Details on the instrumentation and
observing technique are provided by \citet{Agudo2014}.  The data shown
here are measured using the on-off technique with a wobbler of
approximately 45\arcsec.  The data were taken in the framework of the
POLAMI (Polarimetric Monitoring of AGN at Millimeter Wavelengths)
program\footnote{\url{http://polami.iaa.es}} (Agudo et al.~2017a,b, {MNRAS}, subm.;
Thum et al.~2017, {MNRAS}, subm.). The half-power beam widths are
28\arcsec\ and 11\arcsec, respectively. The observations of the
monitoring are spaced on the time-scale of weeks. In this paper we
consider only data from mid 2010 until mid 2014.

\subsection{MOJAVE}
\begin{figure*}
  \centering
  \includegraphics[width=0.07\textwidth,angle=90]{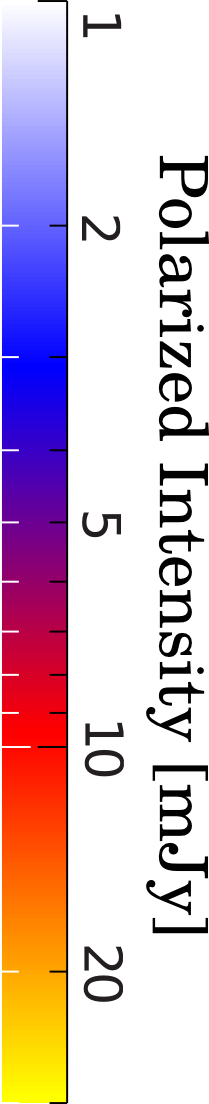}\\
  \begin{minipage}[b][][b]{0.245\textwidth}
    \centering
    \includegraphics[width=\textwidth]{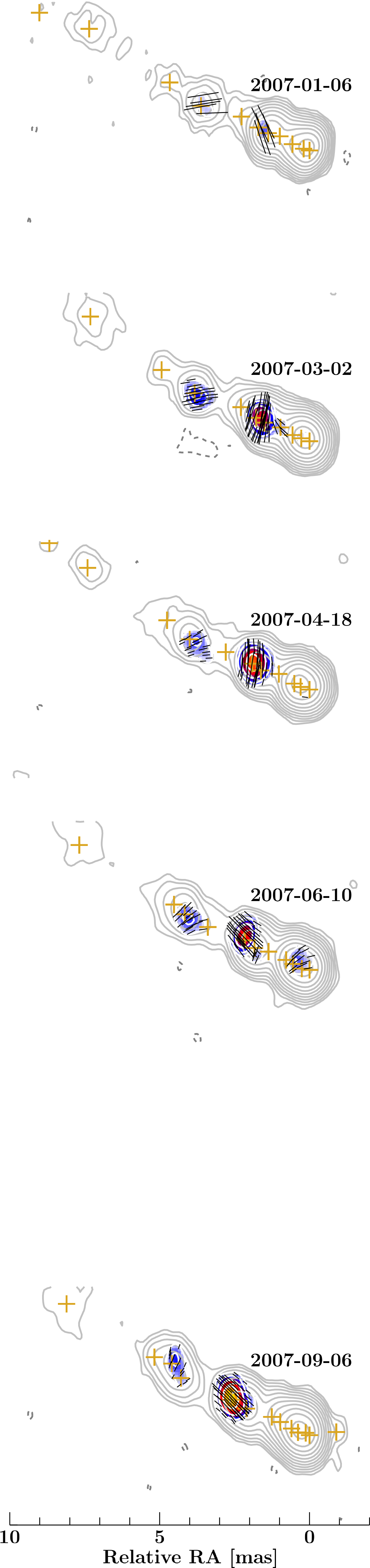}
  \end{minipage}
  \begin{minipage}[b][][b]{0.245\textwidth}
    \centering
    \includegraphics[width=\textwidth]{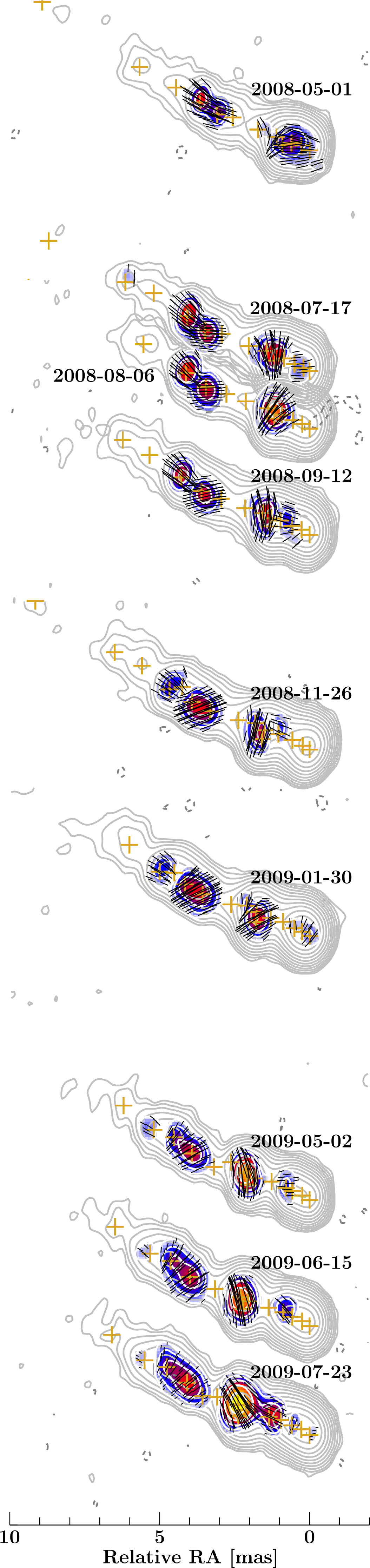}
  \end{minipage}
  \begin{minipage}[b][][b]{0.245\textwidth}
    \centering
    \includegraphics[width=\textwidth]{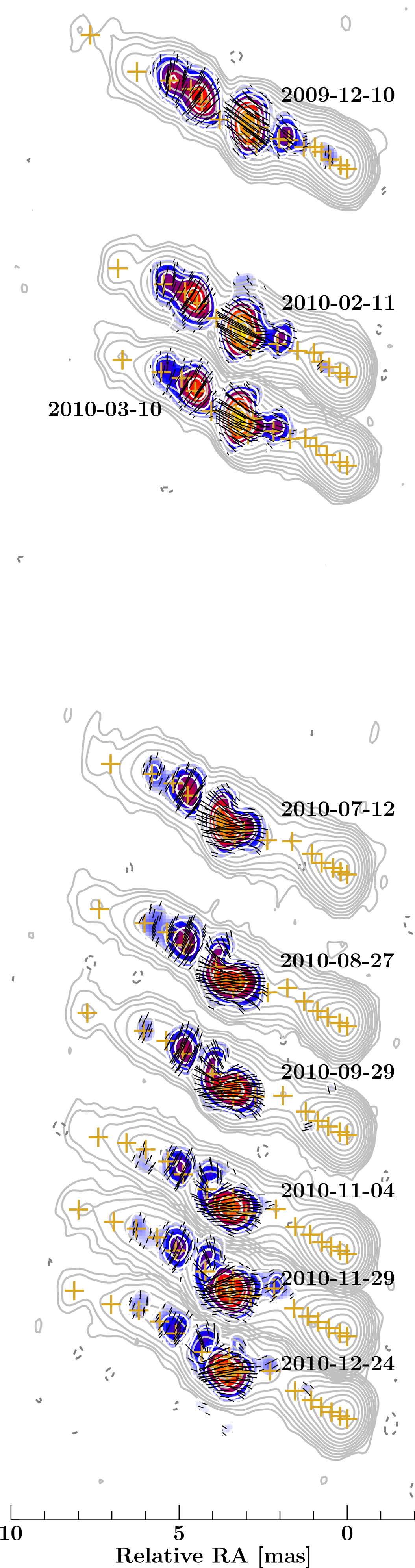}
  \end{minipage}
  \begin{minipage}[b][][b]{0.245\textwidth}
    \centering
    \includegraphics[width=\textwidth]{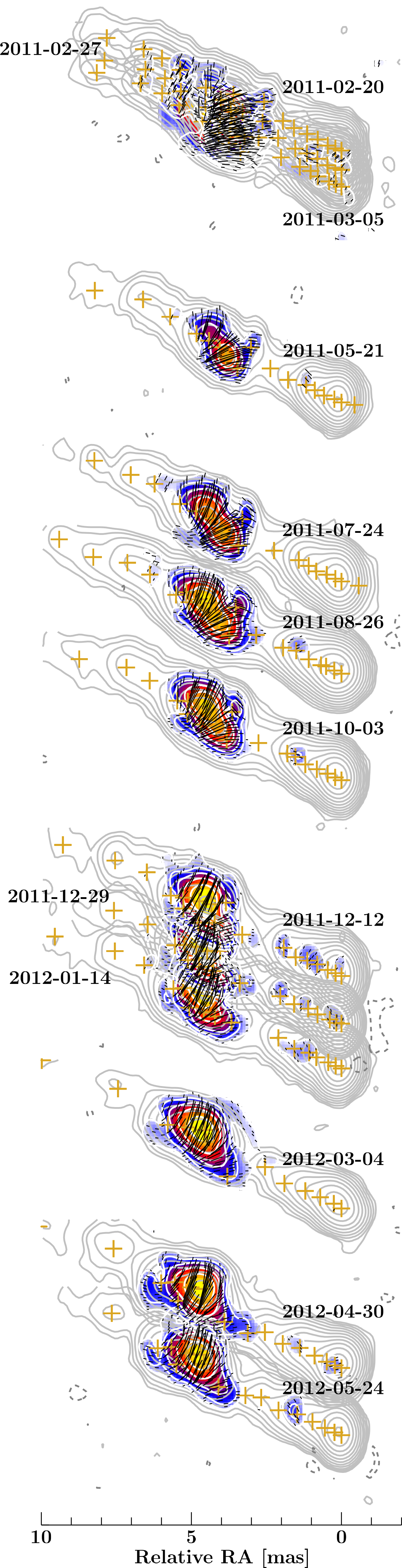}
  \end{minipage}
  \caption{Sequence of 15\,GHz VLBA images from the MOJAVE program
    where we represent total intensity contours ($3\sigma$ above
    background) with overlaid maps of polarized intensity ($5\sigma$
    above background) and corresponding EVPA information drawn as
    vectors on top. The length of the vectors is proportional to the
    polarized intensity. Gaussian model-component positions are
    indicated as orange crosses. The average size of the circular
    model components is $\sim$0.3\,mas for all epochs with a minimum
    size of zero mas at the core.}
  \label{fig:mojave_polflux_maps}
\end{figure*}
\begin{figure*}
\centering
\includegraphics[width=17cm]{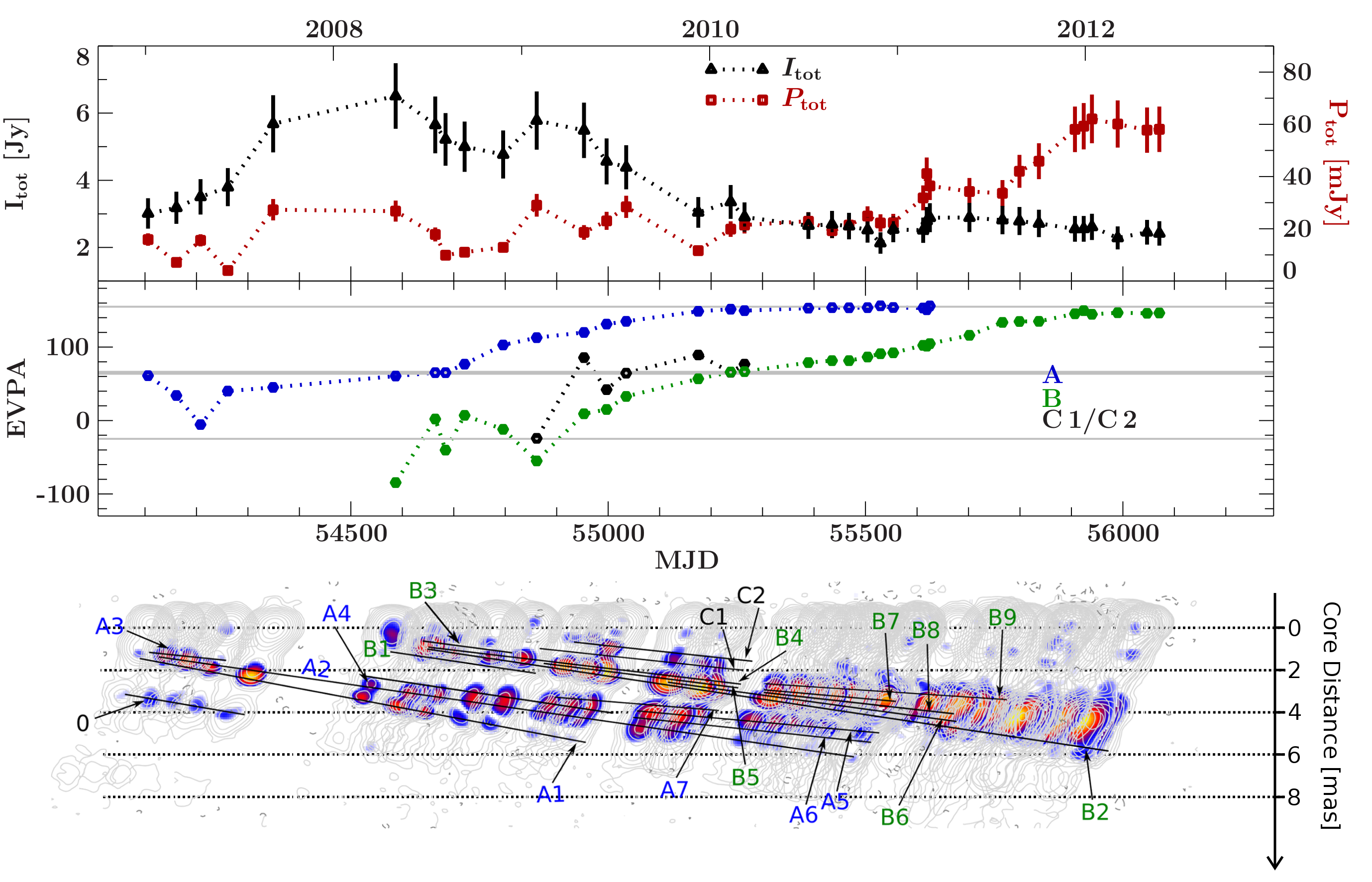}
\caption{\textit{Top panel:} MOJAVE light curve with the total flux
  density (black triangles, for the most significant total flux
  density, $5\sigma$ above the background) and the total polarized
  flux density (dark red squares, integrating the flux densities of
  all model components with at least 1.4\,mJy, which corresponds to
  the average baseline polarized flux density at $3\sigma$ above the
  background); \textit{middle panel:} temporal evolution of the EVPAs
  of the polarized features A (blue), B (green) and C (black). The
  solid gray line at $\sim$63\degr\ corresponds to the jet position
  angle derived from the average angle followed by all tracked
  components. The two thin gray lines at 63\degr-90\degr\ and
  63\degr+90\degr\ mark angles perpendicular to the jet axis. The
  EVPAs are flux-weighted averages for the corresponding feature;
  \textit{bottom panel:} total intensity contours with overlaid
  color-coded maps of polarized intensity as shown in
  Fig.~\ref{fig:mojave_polflux_maps}. Polarized model components
  forming the two main polarized features A and B and their near-ballistic
  trajectories are shown on top of the maps.  Components that
  contribute to the extended most downstream polarized region of
  feature B but where no consistent kinematic model can be found are
  not labeled.}
\label{fig:mojave_lc_map}
\end{figure*}
As part of the MOJAVE monitoring program, 3C~111 has been observed
every few months at 15\,GHz with the VLBA. K08 analyzed the first 17
epochs from 1995 till 2005. We are following up with the upcoming 36
epochs until mid of 2012.  The resulting interferometer visibilities
were calibrated as described in \citet{Lister2005}. We desist from
performing a detailed kinematic analysis and refer to
\citet{Lister2013}, \citet{Grossberger2014} and \citet{Homan2015} in
that regard. The visibilities are naturally weighted and fitted in the
($u$,$v$)-plane using Gaussian model components. The fitted components
are then interpreted with quasi-ballistical trajectories over as many epochs
as possible. See \citet{Grossberger2014} for the details.

We make use of the \texttt{ModelFitPackage} written for the
\texttt{Interactive Spectral Interpretation System} (\texttt{ISIS})
\citep{Houck2000,Grossberger2014}. \texttt{ISIS} is designed for the
spectral analysis of high-resolution X-ray data and provides a
powerful and interactive tool for general astrophysical data analysis
based on the \texttt{S-Lang} scripting language. Its programmability
has turned ISIS into a tool well suited for fitting data of any
kind. In particular, the
\texttt{isisscripts}\footnote{\url{http://www.sternwarte.uni-erlangen.de/isis/}}
provide a pool of functions that are facilitating advanced data
fitting and processing as well as analyzing astrophysical data in
general. The \texttt{ModelFitPackage} interfaces between \texttt{ISIS}
and \texttt{difmap}. That way, we are able to effectively explore the
complex and multi-dimensional $\chi^2$-space and to infer constraints
on the model parameters. For a few prominent sources,
\citet{Grossberger2014} calculates the statistical uncertainties for
model component positions by probing the entire parameter space. When
taking into account parameter degeneracies, we consider a conservative
uncertainty of 0.05\,mas. This value is consistent with the most
probable uncertainty found by \citet{Lister2009}, who analyze the
deviations of the component positions from a common kinematic model
and for the whole MOJAVE sample.
  
For deriving flux densities in the polarized channels Stokes Q and U,
we use the total-intensity Gaussian model components provided by
\citet{Grossberger2014}. We freeze their positions and re-fit only the
flux densities according to \citet{Lister2005}. This approach gives
satisfactory results but we note that in extreme cases of rapid EVPA
changes over core distance, the Q and U visibilities may not be
properly approximated with Gaussian model components. We then
calculate maps of linearly polarized intensity $P=\sqrt{Q^{2}+U^{2}}$
and the $\mathrm{EVPA}=0.5\arctan(U/Q)$ for all 36 epochs treated in
this publication. We show EVPAs uncorrected for Faraday rotation due
to the lack of rotation measure (RM) information covering the entire
parsec-scale jet as observed by MOJAVE. Similar to K08, we assume 15\%
uncertainties on the flux densities. In their appendix,
\citet{Homan2002} empirically derive distinct components to have flux
uncertainties between 5\% and 10\% depending on the component
strength. In our case, the uncertainties may be even higher due to
relatively weak and closely spaced features, justifying the choice of
15\%.

\section{Revealing the polarized jet emission on pc scales}
\label{sec:results}
In recent years, 3C~111 showed three major outbursts, hereafter
labeled A, B, and C, starting in late 2005, 2007, and 2008,
respectively. Figure~\ref{fig:mm_lc} shows light curves taken by IRAM
at 86\,GHz and SMA at 230\,GHz and 350\,GHz. The most prominent
outburst in late 2007 reached a maximum $\gtrsim$13\,Jy at 86\,GHz,
the 2005 and 2008 outbursts reached only about half of that peak flux
density.  The outbursts can be associated with jet activity and the
ejection of apparent superluminal components at 43\,GHz (using
archival data of the Boston University Blazar Group, hereafter
BG\footnote{\url{https://www.bu.edu/blazars/VLBAproject.html}}) and
15\,GHz (MOJAVE). In the following, we study the evolution of the
jet-plasma flow and its polarized emission on parsec scales.

\subsection{Milliarcsecond-scale morphology and evolution}  

\subsubsection{Image analysis}
\label{sec:image_analysis}
Figure~\ref{fig:mojave_polflux_maps} shows the evolution of the jet's
linearly polarized intensity as well as the EVPAs on top of the total
intensity contours at 15\,GHz as a result of MOJAVE (VLBA)
observations.  The images cover the range from 2007-01-06 through
2012-05-24 and are disjunct from the 1999-2006 period studied by
K08. They reveal a number of distinct polarized patterns that can be
described by a comparatively small number of model components close to
the core and become increasingly complex further downstream.  The
polarized patterns are spatially coincident with features in total
intensity that are propagating downstream.  They originate from the
most upstream, stationary feature, the 15\,GHz core that is mainly
unpolarized. In the bottom panel of Fig.~\ref{fig:mojave_lc_map} all
images are ordered along a common time-axis. We identify two distinct
main features that originate in the major outbursts starting in late
2005 and 2007 plus a less dominant feature related to the outburst in
late 2008. These features are hereafter labeled according to their
related outbursts (Fig.~\ref{fig:mm_lc}), namely as features A (blue),
B (green), and C (black) and show persistent polarized emission while
evolving in the downstream direction. The feature labeled with '$0$' is
the remainder of a previous outburst in 2004 (K08) and will therefore
not be discussed in this work.

\begin{figure}
\resizebox{\hsize}{!}{\includegraphics{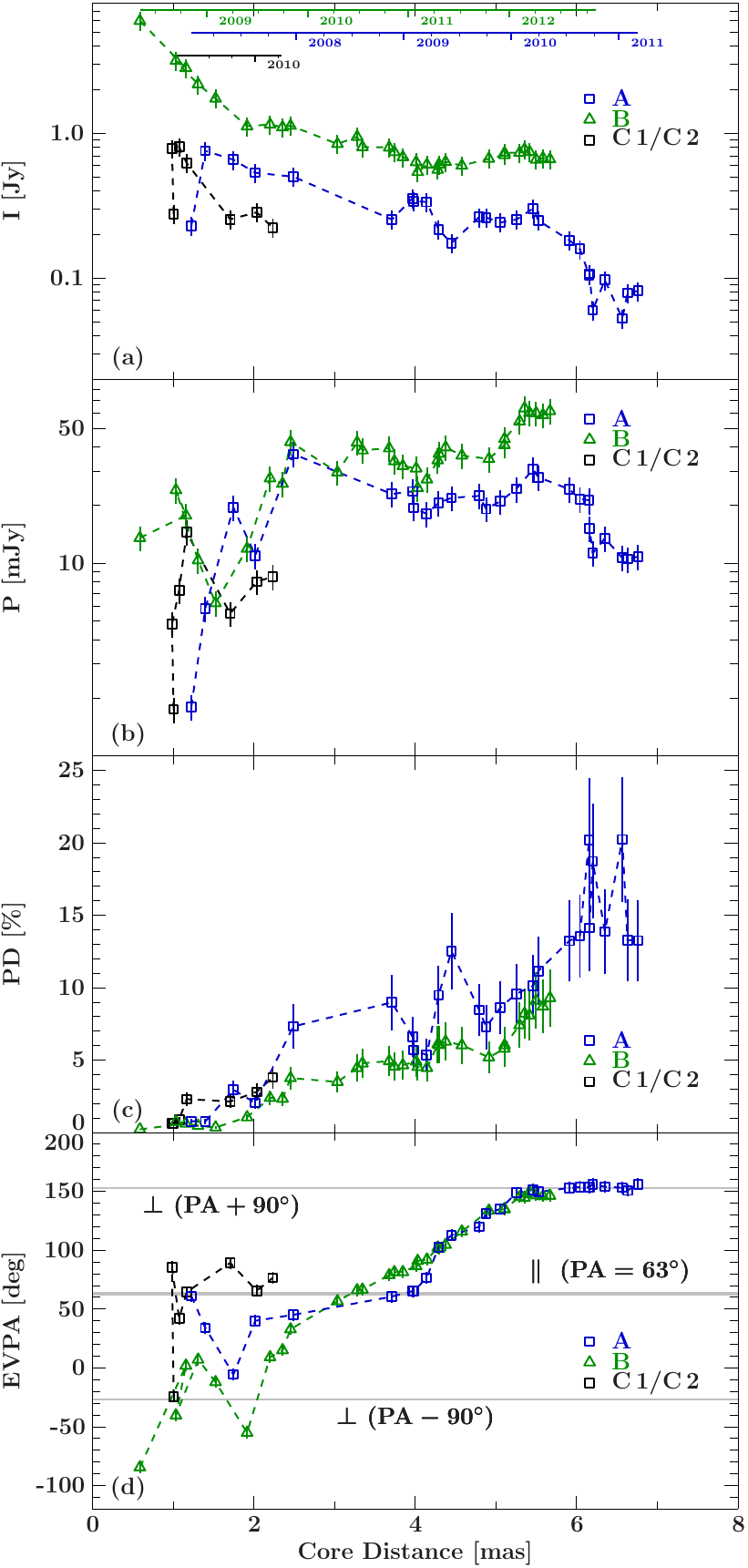}}
\caption{Derived quantities of the total flux density (\textit{panel
      a}), the polarized flux density (\textit{panel b}), the degree
    of polarization (\textit{panel c}) as well as the EVPAs
    (\textit{panel d}) over core distance for the two major polarized
    features A (blue) and B (green) together with the components
    C\,1/C\,2 (black). We use the mean flux-weighted core distance of
    all model components within each feature in a given epoch to
    describe the core distance of that feature. All quantities are
    averaged over the contained model components. The different EVPA
    evolutions are matched by occasional shifts of 180\degr.}
\label{fig:quantities_distance}
\end{figure}

The upper panels of Fig.~\ref{fig:mojave_lc_map} show the epoch-wise
integrated total and polarized flux density. The total intensity light
curve at 15\,GHz as measured by the VLBA peaks in the beginning of
2008 at a level of around 6\,Jy and decreases in flux density towards
2--3\,Jy in the subsequent years.  The two features A and B become
increasingly complex in structure with time and subsequently split
into a number of sub-components (A\,1--A\,7 and B\,1--B\,9).  In
general, their total flux density decreases continuously also with
core distance (Fig.~\ref{fig:quantities_distance}, panel a).  The flux
density of feature A decreases from about 800\,mJy at 1.5\,mas down to
about 60\,mJy at 6--7\,mas, and has two local maxima at $\sim$4\,mas
and $\gtrsim$5\,mas. The feature B is very bright in total intensity
near the core at a level of 6\,Jy and falls rapidly until a distance
of roughly 2\,mas before it reaches a plateau at $\sim$700\,mJy
between 3--6\,mas. A similar behavior was seen for the main features
related to the 1999 outburst by K08.

The evolution of the polarized flux density
(Fig.~\ref{fig:quantities_distance}, panel b) is more complex. Both
features A and B show a common evolution and start off varying
strongly around an average of about 10\,mJy within the inner mas from
the core. The polarized flux density of both features subsequently
increases rapidly to 40\,mJy at 2--3\,mas, and shows a similar
evolution as in total intensity with an overall plateau and local
maxima around 3--6\,mas.  In the end of 2010, the polarized flux
density of feature A fades away and the feature B begins to dominate
the polarized emission reaching a maximum of $\sim$60\,mJy in early
2012.  The leading feature of pattern B describes a local brightening
in polarized intensity between the epochs 2009-05-02 and 2010-03-10,
which is blended with the polarized emission from the other patterns
in the total light curve.  The degree of polarization
(Fig.~\ref{fig:quantities_distance}, panel c) shows less sub-structure
within the uncertainties, but a general increasing downstream trend
starting off with nearly zero percent close to the core up to 15\%
beyond 6\,mas. 

Figure~\ref{fig:quantities_distance} (panel d) shows the EVPAs of the
individual polarized patterns over core distance. In general, the
average EVPAs of the polarized features A, B, and C behave very
similar with distance: a gradual increase of the alignment of their
EVPAs is observed within each feature (see also
Fig.~\ref{fig:mojave_polflux_maps}) and causes the steady downstream
increase of the degree of polarization. The strong EVPA variability
between 1--3\,mas contributes to the low degree of polarization close
to the core due to the partial cancellation of misaligned
EVPAs. Overall, the angles follow a large rotation of about 180\degr\
-- a process that lasts up to four years for each of the spatially
distinct features. The rotation starts around 2--3\,mas at
10--40\degr\ towards being aligned with the jet around 3--4\,mas.  The
alignment of the EVPAs of pattern B with the jet axis happens
coincident with the observed local brightening in polarized intensity.
The swing performed by feature A between 2--4\,mas is slower compared
to feature B. During the brightening, the EVPAs shown in
Fig.~\ref{fig:mojave_polflux_maps} evolve from being overall aligned
within the feature B before epoch 2010-03-10 roughly towards a complex
pattern of EVPAs between 2010-07-12 and 2010-12-24. This leads to some
degree of cancellation of polarization within the beam in the center
of this structure. Beyond 4\,mas from the core, the EVPAs of both
features A and B continue a consistent and smooth rotation of another
90\degr\ towards being transverse at about 150\degr\ during the final
epochs in the end of 2012 featuring regions of up to 6\,mas from the
core.  The averaged EVPAs of pattern C cover a much shorter range in
time and distance but follow a similar behavior as those of the main
patterns A and B.

\begin{figure}
\resizebox{\hsize}{!}{\includegraphics{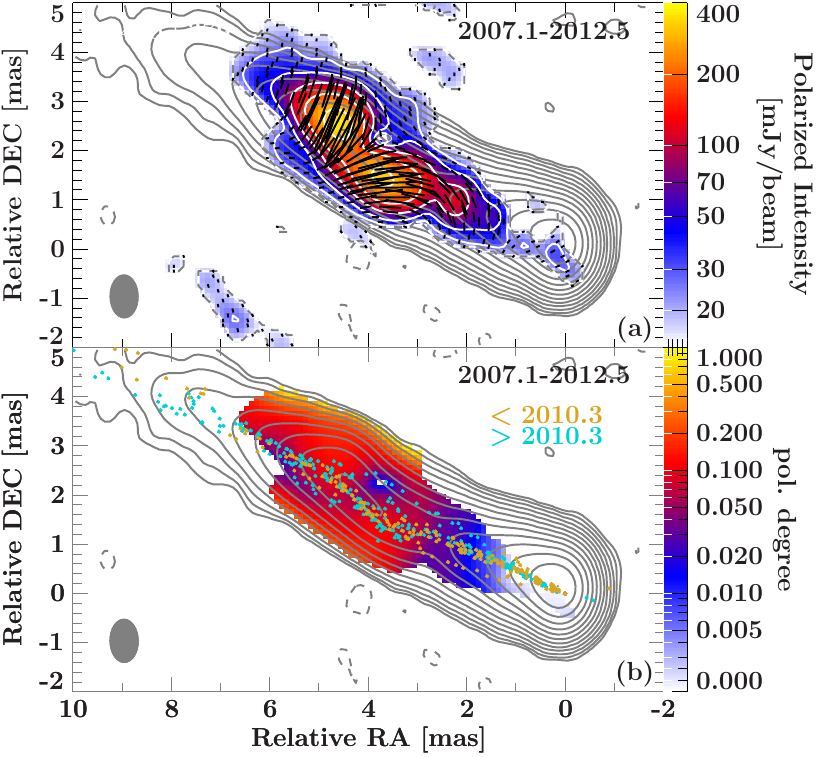}}
\caption{Stacked images with polarimetry information for all involved
  epochs from early 2007.1 until mid 2012.5. We first stack the maps
  of the channels Q and U and then combine those to derive the shown
  maps. \textit{Panel a:} color-coded distribution of the polarized
  intensity with overlaid EVPAs on top of total intensity contours in
  gray. White contours correspond to the polarized
  intensity. \textit{Panel b:} distribution of the degree of
  polarization and overlaid Gaussian (total intensity) model
  components color-coded for components occurring before and after
  2010.3 in orange and blue, respectively.}
\label{fig:stack}
\end{figure}

The apparently similar behavior of the various polarized patterns with
distance along the jet motivates an inspection of the stacked image of
all individual observations between 2007.1 and 2012.5 as shown in
Fig.~\ref{fig:stack}. In such a stacked image of multiple observations
of a jet with moving features, one expects effective depolarization
unless the polarized emission of different features as a function of
distance along the jet is strongly correlated.  The stacked image
shows indeed strong polarization with its EVPAs tracing a continuous
rotation between 2--6\,mas from the jet core accompanied by an
increase of the net polarized intensity and the degree of
polarization.  The distribution of the (total intensity) model
components (Fig.~\ref{fig:stack}, panel b) appears as overall straight
but tentatively traces a bend towards the south near a
low-polarization region. The figure also shows that the edges of the
jet flow become illuminated by a couple of jet components at different
times.
\begin{table*}
  \caption{List of all polarized model components as part of the polarized features A and B with corresponding proper motions in units of mas/yr. The leading components A\,1/A\,2 and B\,1/B\,2 are followed by the trailing components on the bottom. See also the Figs.~\ref{fig:xy_A_appendix} and \ref{fig:xy_B_appendix} in Appendix C for fits of the PAs followed by the jet components in  \textit{x/y}-space. For A\,2, no PA is given for the range 3\,mas$-$4.5\,mas due to the insufficient number of model components.}
  \label{tab:components}
  \vspace{0.5cm}
  \centering
  \begin{tabular}{lr@{$\pm$}lr@{$\pm$}llr@{$\pm$}lr@{$\pm$}l}
    \hline\hline 
    A & \multicolumn{2}{c}{PA~[deg]} & \multicolumn{2}{c}{$\mu$~[mas/yr]}  & B & \multicolumn{2}{c}{PA~[deg]}  & \multicolumn{2}{c}{$\mu$~[mas/yr]} \\
    \hline
    A\,1                                   & 66.0 & 0.4      & 1.731& 0.016  & B\,1 & 62.54&1.13  & 1.67& 0.06 \\ 
    A\,2 ($<3\,\mathrm{mas}$)              & 67.9 & 0.6      & 1.30& 0.14    & B\,2 & 63.6&0.3  & 1.483& 0.009 \\ 
    A\,2 ($3\,\mathrm{mas}-4.5\,\rm{mas}$) & \multicolumn{2}{c}{}  & 1.56 & 0.09    & B\,3 & 67.6&2.3  & 0.80& 0.12 \\ 
                      A\,2 ($>4.5\,\mathrm{mas}$)            & 57.9 & 0.3    & 1.28& 0.02  & B\,4 & 69.6&0.7  & 1.14& 0.04 \\
                      A\,3                                   & 70.7 & 2.3      & 0.91& 0.16    & B\,5 & 55.56&4.17  & 1.521& 0.014 \\
                      A\,4                                   & 68.3 & 0.5      & 1.40& 0.04    & B\,6 & 53.3&0.4  & 1.18& 0.03 \\
                      A\,5                                   & 59.3 & 0.4      & 0.97& 0.04    & B\,7 & 41.7&0.8 & 1.14& 0.08 \\
                      A\,6                                   & 62.8 & 0.3    & 1.15& 0.04    & B\,8 & 52.7&0.5  & 0.77& 0.06 \\
                      A\,7                                   & 69.0 & 0.5      & 1.325& 0.017  & B\,9 & 80.0&0.5  & 1.12& 0.05 \\
                                                            &     \multicolumn{2}{c}{}             &   \multicolumn{2}{c}{}               & C\,1 & 66.3&1.2  & 1.20& 0.05 \\
                                                             &    \multicolumn{2}{c}{}              &   \multicolumn{2}{c}{}               & C\,2 & 65.7&1.8  & 1.09& 0.07 \\
    \hline
  \end{tabular}
\end{table*}
\begin{figure}
\resizebox{\hsize}{!}{\includegraphics{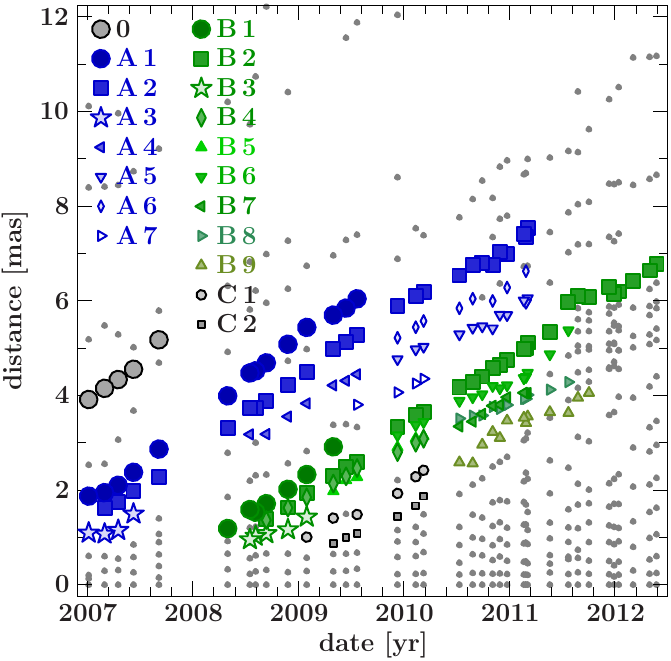}}
\caption{Distances over time for all Gaussian model components required to fit
  the total intensity visibility data of all analyzed MOJAVE epochs
  (gray). In color, we only show model components describing the
  polarized features A (blue) and B (green), if their polarized flux
  density exceeds the given threshold of 1.4\,mJy ($3\sigma$). In
  cases of components that are tracked over more than four epochs, we set a
  threshold of 1.0\,mJy ($2\sigma$). The two leading components A\,2
  and B\,2 are highlighted with large, filled squares. The two
  components A\,3 and B\,3 that form in the wake of these leading
  components during the first epochs of the features A and B, but fade
  quickly, are denoted as stars. We do not trace in detail the
  evolution of components in the wake of B\,2 after 2011.4 due to the
  increased complexity of the polarized brightness distribution that
  cannot entirely be described by ballistic Gaussian model
  components.}
\label{fig:date_dist}
\end{figure}

\subsubsection{Kinematic analysis}
\label{sec:image_analysis}
We can describe the dynamics of the polarized features A, B, and C
quantitatively by \textsc{i}) modeling the
total intensity emission with a small number of Gaussian components
(following \citealt{Grossberger2014}), \textsc{ii}) measuring the
linear polarization of these Gaussian components, and \textsc{iii})
performing a kinematic study of components with significant
polarization. For the latter, we constrain ourselves to components
with a polarized flux density of at least 1.4\,mJy, which, on average,
corresponds to a $3\sigma$ detection with respect to the
background. We reduce this lower threshold to 1\,mJy, i.e., $2\sigma$
only in case of model components that can be tracked over more than
four epochs.

\begin{figure*}
  \centering
  \includegraphics[width=0.48\textwidth]{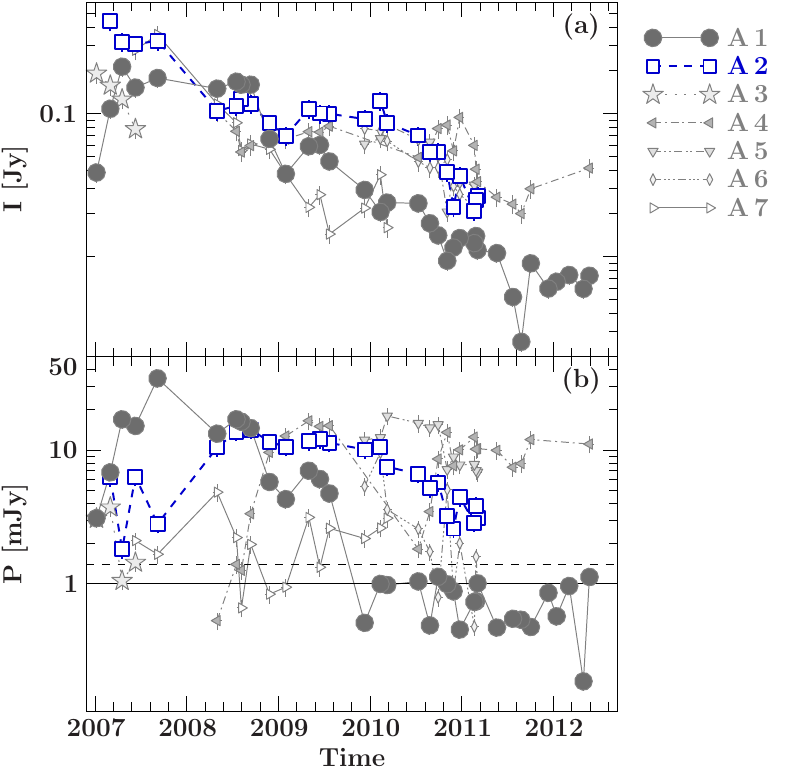}\hfill
  \includegraphics[width=0.48\textwidth]{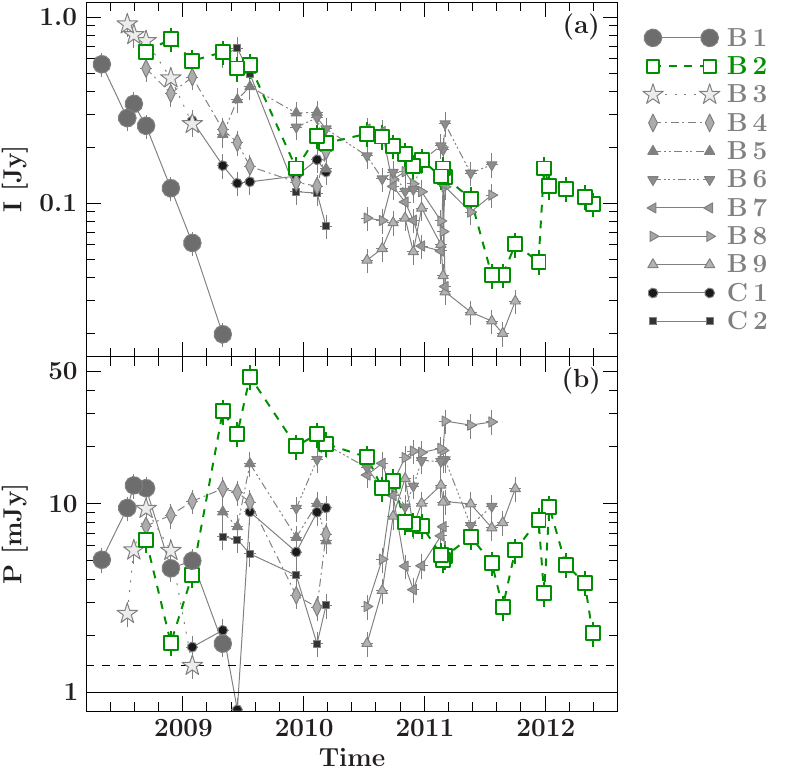}
  \caption{Evolution of the total (\textit{panels a}) and polarized flux
    density (\textit{panels b}) for selected components of group A and
    B. For all components the total flux decreases with
    time. Components identified as trailing components are held in
    gray. The $2\sigma$ ($3\sigma$) thresholds for polarized
    components to be considered in Fig.~\ref{fig:quantities_distance}
    and Fig.~\ref{fig:date_dist} are denoted as solid (dashed) black
    lines.}
  \label{fig:compfluxAB}
\end{figure*}

Most of these components follow near-ballistic trajectories with similar
PAs (Fig.~\ref{fig:mojave_lc_map}). Table~\ref{tab:components} lists
the polarized components with inferred proper motions on the sky.
The distances of the selected polarized components and the unpolarized
jet model components as a function of time are shown in
Fig.~\ref{fig:date_dist}.  The leading components of the features A
and B (A\,1/A\,2 and B\,1/B\,2, respectively) can be tracked over
almost the full time range, and dominate both the total and polarized
flux densities of the parsec-scale jet at nearly all time
(Fig.~\ref{fig:compfluxAB})\footnote{We exclude the polarized fluxes
  before 2009.2 of component A\,7 for further calculations due to their
  questionable detectability.}.  The total flux density evolutions of
the components A\,1 and B\,1 are considerably different. A\,1 can be
tracked over all 36 epochs and only slowly decreases in flux. The
component B\,1, in contrast, fades within seven epochs. The polarized
flux density of both A\,1 and A\,2, as well as B\,2, rapidly increase
during their first epochs, which gives rise to the observed peaks in
the integrated polarization light curve
Fig.~\ref{fig:quantities_distance} (panel b) at around 2--3\,mas.  The
components A\,3 to A\,7 and B\,3 to B\,11 form behind the leading
components of both polarized features and their flux densities are
overall lower. We therefore identify them as trailing components
\citep{Agudo2001} in the wake of the leading components similar to
those observed by K08 after the 1996 outburst of 3C~111.

The leading components A\,1 and B\,1 are both found on ballistic
trajectories in $(x/y)$-space with position angles (PAs) of about
66\degr\ and 63\degr, and show comparable proper motions of
$1.731\pm 0.016\,\mathrm{mas/yr}$ and $1.67\pm 0.06\,\mathrm{mas/yr}$,
respectively (see also Figs.~\ref{fig:xy_A_appendix} and \ref{fig:xy_B_appendix} in
Appendix C). This is again similar to the speed of the leading component
associated with the 1996 outburst (see K08). The components A\,2 and
B\,2 follow PAs of about 68\degr\ and 64\degr, respectively. The PA of
A\,2 changes to about 58\degr\ after 3--4\,mas. While we find a proper
motion of $1.48\pm 0.01\,$mas/yr for B\,2, A\,2 cannot be
described ballistically but shows signs for moderate acceleration in
longitudinal direction from
$1.30\pm 0.14$\,mas/yr ($<3$\,mas) to
$1.56\pm 0.09$\,mas/yr (3$--$4.5\,mas) before
decelerating again to
$1.28\pm 0.02$\,mas/yr ($<3$\,mas).
A similar behavior may be inherent to the component A\,1, despite
being statistically insignificant. We therefore remain with a
ballistical description of the same.

\begin{figure}
\resizebox{\hsize}{!}{\includegraphics{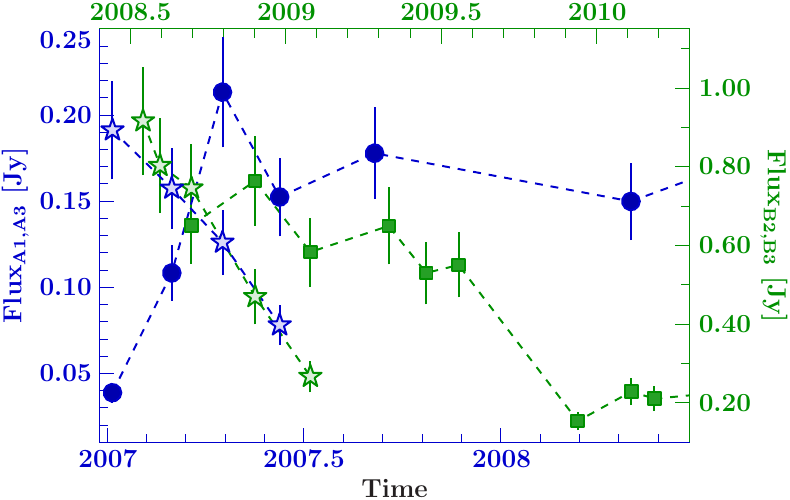}}
\caption{Flux density evolution of the short-lived components A\,3
  (blue stars) and B\,3 (green stars) together with the leading
  components A\,1 (blue circles) and B\,2 (green circles). Due to
  their different appearance times, the time axes of both components
  are different.}
\label{fig:flux_rarefactions}
\end{figure}

In both polarized features (A and B), we find components with slower
proper motions being formed early on (A\,3:
$0.91\pm 0.16$\,mas/yr; B\,3:
$0.80\pm 0.12$\,mas/yr), both rapidly decreasing in flux
density (see Fig.~\ref{fig:flux_rarefactions}).  The component A\,3
loses more than half of its flux density from $\sim$200\,mJy to
$\sim$80\,mJy within less than half a year, while B\,3 shows a more
drastic decrease in flux density of a factor of five from $\sim$900\,mJy to
$\sim$250\,mJy over four months. This is reminiscent of the similar
behavior of component ``F'' in K08. The total flux density of both
A\,3 and B\,3 starts off larger than that of the leading components
A\,1 and B\,2, which increase in flux density during the decreasing evolution
of A\,3 and B\,3.

In general, all components in the wake of A\,1/A\,2 and B\,1/B\,2 have
lower proper motions than the leading components (between
0.8--1.4\,mas/yr). Also, their inherent amount of variability both in
total and polarized flux density is overall larger. Amongst these
trailing components, the component A\,4 forms just after 3\,mas from
the core with a proper motion of $1.40\pm 0.04$\,mas/yr and
follows a trajectory with a PA of $68.3\pm 0.5$\degr. It subsequently
appears to split into A\,5 and A\,6 at around 4.5\,mas from the core,
which continue along different PAs of $59.3\pm 0.4$\degr\ (A\,5) and
$62.8\pm 0.3$\degr\ (A\,6) at smaller velocities of
about 1.0\,mas/yr. If the components A\,4, A\,5 and A\,6 would
describe the same underlying perturbation, we could put them into
context with component A\,2, which shows decelerating behavior
downstream of approximately 4.5\,mas. Their changing PAs, however,
challenge this interpretation.

\begin{figure}
  \resizebox{\hsize}{!}{\includegraphics{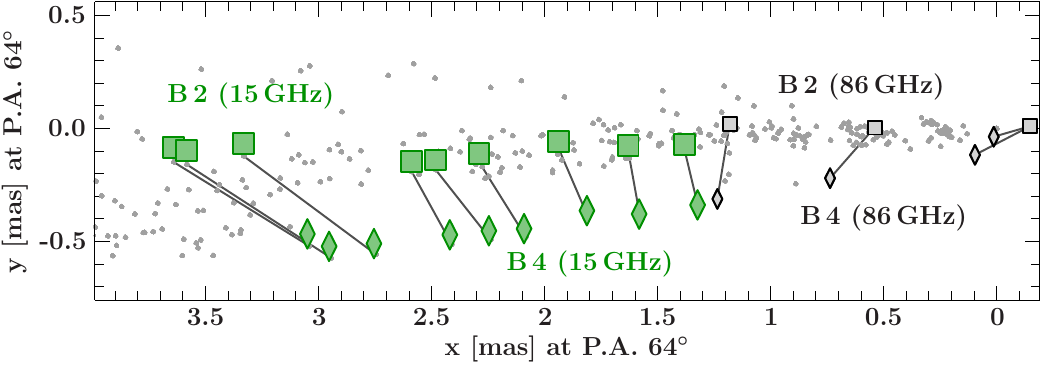}}
  \caption{($x,y$)-plot emphasizing the components B\,2 (triangles)
    and B\,4 (diamonds) projected on a PA of 64\degr as observed at
    15\,GHz. The positions of all other model components at 15\,GHz
    are drawn in gray. Gray, solid lines connect the positions of B\,2
    and B\,4 at the same epochs. Gray symbols resemble model
    components found at 86\,GHz further upstream (Schulz, priv. comm.)
    that we relate with the components B\,2 and B\,4 at 15\,GHz.}
  \label{fig:64a70a}
\end{figure}

The polarized feature B describes the most stable polarized region
that becomes increasingly complex in later epochs, eventually
dominating the polarized intensity of the entire jet.  The components
B\,2 and B\,4, seem to emerge out of a single unresolved component
with an initial flux density of $1.15$\,Jy in epoch 2008.6, with B\,2
becoming the more dominant and faster component in subsequent epochs
(B2: $\sim$1.5\,mas/yr; B\,4: $\sim$1.1\,mas/yr).
Figure~\ref{fig:64a70a} shows the trajectories of both these
components. The figure also includes two components that are observed
and identified on the course of the 2007 outburst with help of
higher-resolution observations with the Global Millimetre VLBI array
by Schulz et al. (in prep.). The components B\,2 and B\,4 at 15\,GHz
appear as a continuation of these components at 86\,GHz. Their
velocity difference might reflect a shear within the plasma that we
observe as feature B (cf. Discussion).

\subsection{Analysis of the brightness temperature distribution}
\label{sec:tb}
\begin{figure*}
\sidecaption
\includegraphics[width=12cm]{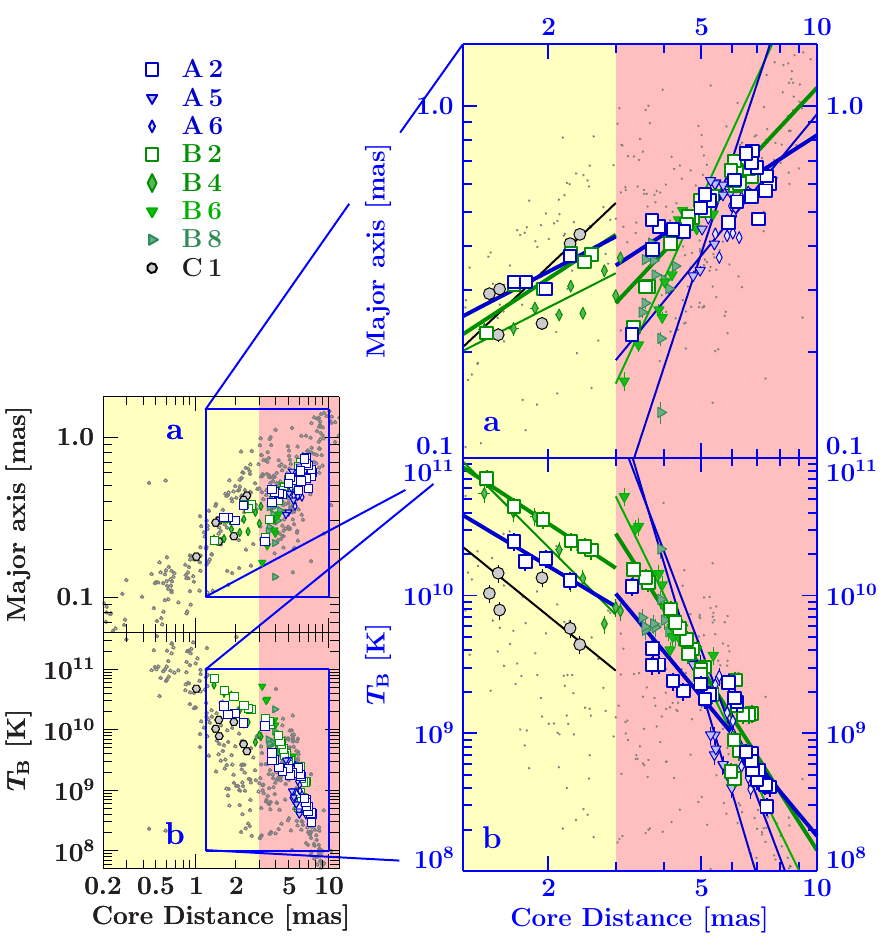}
\caption{Brightness temperature and major axis size against the core
  distance for all model components of all epochs (gray). The colored
  symbols correspond to model components that can be associated with
  the polarized features A and B. The components A\,2 and B\,2 probe
  the transition region at around 3\,mas and are therefore emphasized
  with enlarged squares of white filling. Lines correspond to linear
  regression fits of data up- and downstream of 3\,mas.}
\label{fig:tb_kinematics}
\end{figure*}
We have shown above that the evolution of the EVPAs of both features A
and B behave similarly with time and core distance. The consistent
orientation parallel to the jet around 3--4\,mas from the core and the
foregoing drastic increase in polarized flux density, lead us to
further study this region. Figure~\ref{fig:tb_kinematics} shows the
dependence of the brightness temperature $T_\mathrm{B}$ and the sizes
of the circular Gaussian model components $d$ over the core distance
$r$.

The brightness temperature has been shown to follow a power law over
core distance for samples of AGN jets
\citep{Kadler2005,Pushkarev2012}, for the particular case of 3C~111
(K08) and for other individual sources \citep[e.g., NGC~1052:
][or S4~1030$+$61: \citealt{Kravchenko2016}]{Kadler2004}. We note, however, that the
RadioAstron Space VLBI resolution allows us to come even closer to the
apparent jet base and delivers higher brightness temperatures as
expected \citep[e.g.,][]{Gomez2016,Kovalev2016}. If the magnetic field
$B \propto r^b$, the particle density $N \propto r^n$ and the jet
diameter $d \propto r^l$ evolve like power laws with distance $r$ from
the core, the brightness temperature can be described as
$T_\mathrm{B}\propto r^{s}$ \citep[][K08]{BK1979}. The power-law index s (with $s<0$)
can then be expanded as
\begin{equation}
  \label{eq:plindex}
  s=l+n+b\,(1-\alpha) \quad,
\end{equation}
where $\alpha$ is the spectral index, characterizing the flux-density
spectrum via $S_\nu \sim \nu^{-\alpha}$.

Figure~\ref{fig:tb_kinematics} makes clear that this simplified ansatz
can successfully describe the measured brightness temperature and jet
diameter.  We adopt uncertainties of 0.05\,mas for $r$, 15\% on
$T_\mathrm{B}$ and 0.01\,mas on $d$, which corresponds to the scatter
of all measured major axes. A sudden decrease of the model component
sizes is apparent at a distance of around 3\,mas from the core. This
jump is accompanied by a jump in $T_\mathrm{B}$ at that distance for
the two components A\,2 and B\,2 that probe this transition region.
We find the size of the component A\,2 to decrease from
$0.43\pm 0.03$\,mas to $0.352\pm 0.004$\,mas and for
component B\,2 from $0.43\pm 0.02$\,mas to
$0.275\pm 0.003$\,mas when extrapolating the measured
power-laws to the discontinuity at around 3\,mas. The extrapolated
brightness temperatures increase from
$8_{-3}^{+4}\times 10^{9}$\,K to
$1.03\pm 0.09 \times 10^{10}$\,K for A\,2 and from
$1.6_{-0.4}^{+0.5}\times 10^{10}$\,K to
$ 2.8_{-0.2}^{+0.3}\times 10^{10}$\,K for B\,2. For constant
flux, two measurements of the brightness temperature, i.e.,
$T_{\mathrm{B},1}$ and $T_{\mathrm{B},2}$, are related to the
corresponding component sizes $d_1$ and $d_2$ by
$T_{\mathrm{B},1}/T_{\mathrm{B},2}\propto (d_{2}/d_{1})^{2}$. We find
$T_{\mathrm{B},1}/T_{\mathrm{B},2}=0.8\pm 0.3$ and
$(d_{2}/d_{1})^{2}=0.82\pm 0.11$ for component A\,2 as well as
$T_{\mathrm{B},1}/T_{\mathrm{B},2}=0.57\pm 0.16$ and
$(d_{2}/d_{1})^{2}=0.6\pm 0.3$ for B\,2. Within the uncertainties,
the sudden decrease of the component sizes is consistent with the
increase of $T_\mathrm{B}$ for adiabatic knots. The other plotted
components have an insufficient number of traceable counterparts
upstream or downstream of 3\,mas and do not add further information to
these results.
 
In Table~\ref{tab:slopes} we list the measured indices $l$ and ($s-l$)
upstream and downstream of 3\,mas. We exclude component B\,8 from the
fits. It describes an unrealistically steep power law, probably as
result of its strongly variable flux density.
\begin{table}
  \caption{Slopes for a power-law relation between the size of 
    the major axis, $d$, and the brightness temperature 
    $T_\mathrm{B}$ against the core distance $r$  
    ($\propto r^l$ and $\propto r^s$) in 
    log--log space. Data up- and downstream of a distance of 
    around 3\,mas are fitted with separate power laws. 
    The component B\,8 is excluded from the fits due to unrealistically 
    steep slopes suggested by the data.}
  \vspace{0.5cm}
  \centering
  \resizebox{\columnwidth}{!}{
  \begin{tabular}{lr@{$\pm$}lr@{$\pm$}lr@{$\pm$}lr@{$\pm$}l}
    \hline\hline 
     & \multicolumn{2}{c}{$l_{<\,3\,\mathrm{mas}}$} & \multicolumn{2}{c}{$l_{>\,3\,\mathrm{mas}}$} & \multicolumn{2}{c}{$s-l_{<\,3\,\mathrm{mas}}$} & \multicolumn{2}{c}{$s-l_{>\,3\,\mathrm{mas}}$}\\
    \hline
    A\,2 & 0.57& 0.04 & 0.709& 0.004 & -2.2& 0.2 & -4.07& 0.03\\
    A\,5 &  \multicolumn{2}{c}{\ldots}& 3.330& 0.010 &  \multicolumn{2}{c}{\ldots}& -12.73& 0.05\\
    A\,6 &  \multicolumn{2}{c}{\ldots}& 1.339& 0.006 &  \multicolumn{2}{c}{\ldots}& -8.71& 0.04 \\
    B\,2 & 0.71& 0.03 & 1.172& 0.004 & -2.55& 0.15 & -5.57& 0.03\\
    B\,4 & 0.56& 0.02 &  \multicolumn{2}{c}{\ldots}& -3.35& 0.11 &\multicolumn{2}{c}{\ldots} \\ 
    B\,6 &  \multicolumn{2}{c}{\ldots}& 2.380& 0.010 &  \multicolumn{2}{c}{\ldots}& -8.08& 0.06 \\
    C\,1 &  \multicolumn{2}{c}{\ldots}& 1.03& 0.04 &  \multicolumn{2}{c}{\ldots}& -3.29& 0.18 \\
    \hline
  \end{tabular}
}
  \label{tab:slopes}
\end{table}
Compared to a free expansion ($l=1$), we find reduced expansion rates
upstream of 3\,mas. Beyond 3\,mas, however, we find several components
with indices as high as 1--3, averaging $\sim$1.7. The index
combination ($s-l$) is a measure of the gradients of the magnetic
field and the gas density. It shows moderate values upstream of 3\,mas
and extremely steep values beyond.

\begin{figure}
  \centering
  \resizebox{\hsize}{!}{\includegraphics{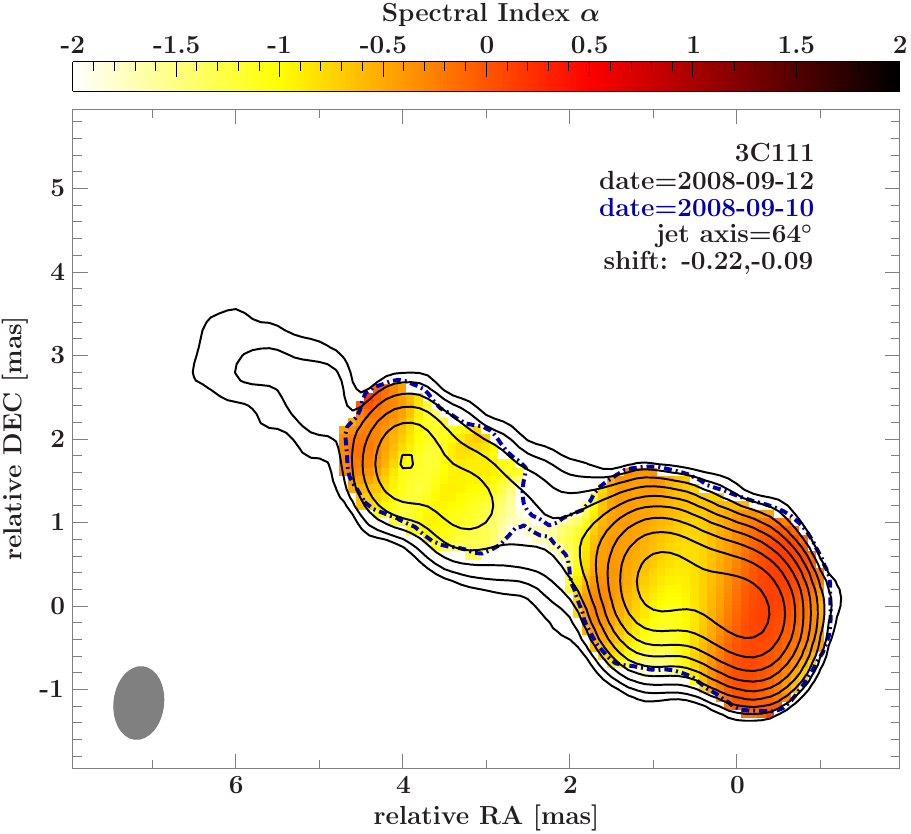}}
  \caption{Spectral index maps for quasi-simultaneous observations at
    43\,GHz (BG, epoch 2008-09-10) and 15\,GHz (MOJAVE,
    2008-09-12). Both maps are restored with a common beam shown on
    the bottom left and the spectral index $\alpha =
    \log{S_1/S_2}/\log{\nu_2/\nu_1}$ is computed accordingly for each
    pixel. The required shift of the 43\,GHz map relative to the one
    at 15\,GHz is $-0.22$\,mas in right ascension and $-0.09$\,mas in
    declination. The shift is determined by matching the flux-averaged
    mean $x$/$y$ positions of the brightest components in both
    individual maps, excluding the core.}
  \label{fig:spixmap}
\end{figure}

To characterize the influence of $\alpha$ on $s$, we calculate
spectral-index maps between 15\,GHz and 43\,GHz (BG data).  We chose
four separate epochs during which jet plasma components occupy the
region beyond 3\,mas and for which closely
separated\footnote{The separation is given by 2, 3, 4, and 6\,days for
  the 43\,GHz (15\,GHz) epochs 2008-09-10 (2008-09-12), 2011-07-24
  (2011-07-21), 2011-02-27 (2011-03-01), and 2009-01-30.}
observations at 15\,GHz and 43\,GHz are available.  We compare
flux-weighted positions of optically thin regions of the total
intensity maps to determine a core shift of $RA=-0.22$ and $DEC=-0.09$
between the two maps \citep[cf., e.g.,][]{Kadler2004}.  Both maps are
restored with a common beam enclosing the two single beams at 15\,GHz
and 43\,GHz.  The exemplary map in Fig.~\ref{fig:spixmap} reveals a
pronounced gradient from an optically thick core with a flat power-law
spectrum towards optically thin jet emission with $\alpha \sim -1$
downstream of 3\,mas -- a behavior that is observed in all four
analyzed spectral index maps at different times. As a cross-check, we
extract the integrated flux densities from the emission region
downstream of 3\,mas for both the restored 15\,GHz and 43\,GHz
map and calculate the averaged spectral index using the flux-density
ratios. We derive consistent values of $\alpha \sim -1$ for all
four tested epochs.

The parameter $b$ describes the geometry of the magnetic field and
cannot be directly measured with our data.  In the idealized cases of
a pure toroidal field, a value of $b=-1$ would apply. Similarly,
$b=-2$ would describe a pure axial field.  For these two cases, we can
use the measurements of $s-l$ and $\alpha$ to constrain the gradient
of the particle density along the jet.  We consistently find power
laws $r^n$ with the index steepening at the distance of 3\,mas from
the core from $n_{<3\,\mathrm{mas}}\sim -0.2\,(1.8)$ to
$n_{>3\,\mathrm{mas}}\sim -2.1\,(-0.1)$ for component A\,2 and from
$n_{<3\,\mathrm{mas}}\sim -0.6\,(1.4)$ to
$n_{>3\,\mathrm{mas}}\sim -3.6\,(-1.6)$ for component B\,2. The
numbers consider a magnetic field with $b=-1$ ($b=-2$).

\section{Discussion}
\label{sec:discussion}
We have reported on the evolution of two features A and B both strong
in total and polarized intensity through the VLBI jet in 3C~111. Their
evolution follows a similar pattern for both groups of components
between 2--6\,mas, which is largely consistent with results from K08
and an independent study by \citet{Homan2015}, who suggest a very
similar velocity pattern with signs for accelerated motion upstream of
3--4\,mas and decelerated motion beyond.  In the following we discuss
our results and describe possible scenarios that could explain these
observed jet features, which most likely reflect shocked
plasma. Therefore, the observed and derived jet-intrinsic quantities
can not be interpreted as those describing an unperturbed flow.

\subsection{Inner 2\,mas from the core}
Schulz et al. (in prep.) report on very-high angular resolution GMVA
observations of the inner 2\,mas of the jet of 3C\,111 during the
outburst that has led to the ejection of the components associated
with feature B.  In Fig.~\ref{fig:64a70a}, we show the positions of
the 86\,GHz model components that can be associated with the 15\,GHz
jet components B2 and B4 on the basis of their temporal evolution at
both frequencies. The 86\,GHz images show a highly complex structure
and dynamical evolution of the individual jet components on these
small scales. 

\subsubsection{Possible effects of Faraday rotation}
\begin{figure}
  \centering
  \resizebox{\hsize}{!}{\includegraphics{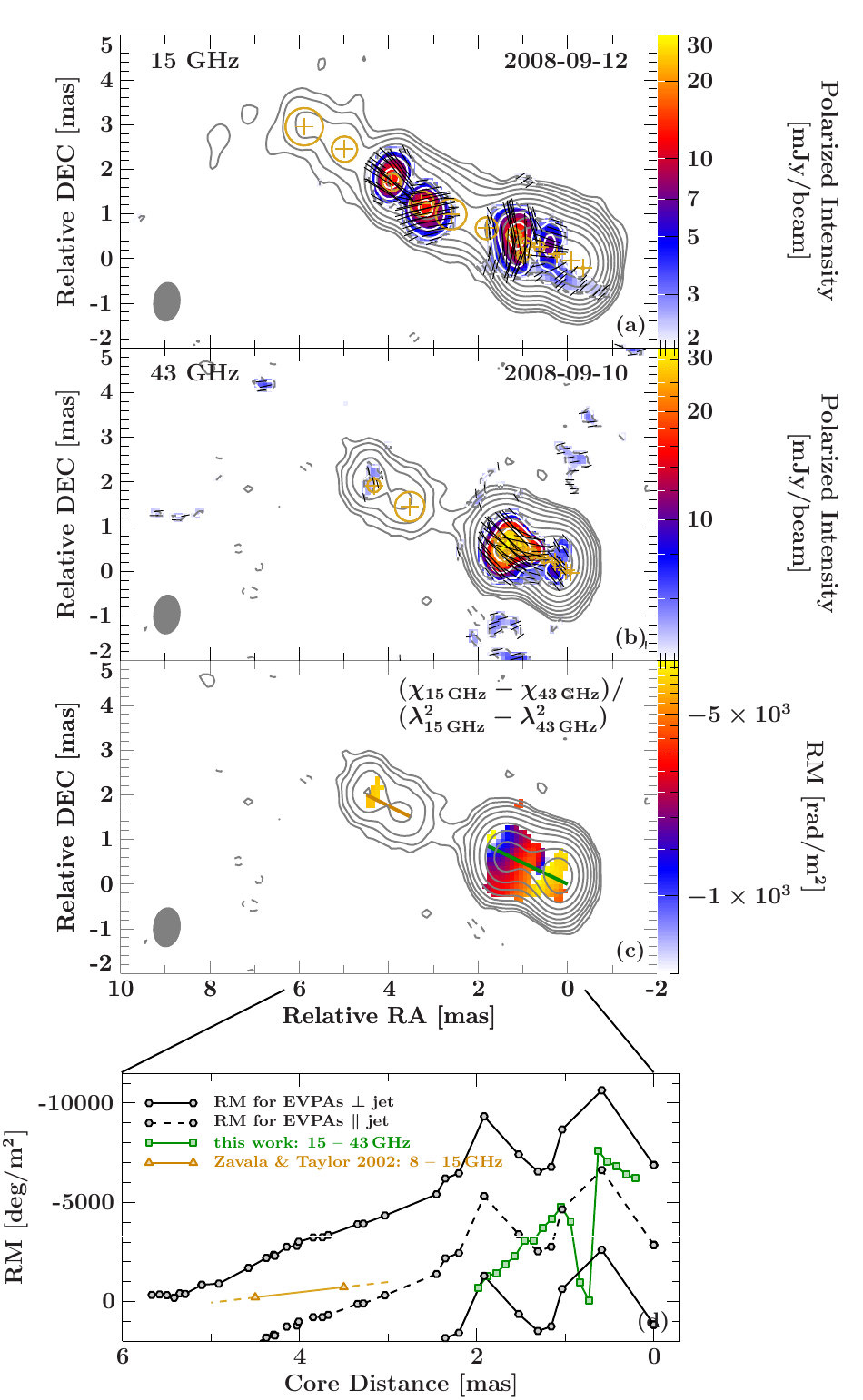}}
  \caption{RM map (\textit{panel c}) between two adjacent maps at
    15\,GHz/MOJAVE (\textit{panel a}) and 43\,GHz/BG (\textit{panel
      b}) that have also been used to calculate the spectral index in
    Fig.~\ref{fig:spixmap}. The polarized emission is overlaid in
    color on top of the total intensity contours, both at a baseline
    intensity of 3$\sigma$. We only calculate values of RM where we
    detect polarized emission $>3\sigma$ both at 15\,GHz and
    43\,GHz. We apply the same relative core shift between the maps at
    both frequencies as estimated for Fig.~\ref{fig:spixmap} and the
    identical envelope beam. \textit{Panel d} shows RM cuts along the
    indicated ridge lines for our measurement upstream of 2\,mas
    (green squares) and for a measurement by \citet{Zavala2002}
    between 3.5\,mas and 4.5\,mas (orange triangles). We also show as
    black solid and dashed lines the RM distribution along the jet
    that is required to explain the observed EVPA evolution of feature
    B (Fig.~\ref{fig:quantities_distance}) with respect to either an
    intrinsically perpendicular or parallel field, respectively. The
    two RM distributions shown for an intrinsically perpendicular
    field reflect the ambiguity of $\pi$.}
  \label{fig:rm}
\end{figure}

The unresolved knots that emerge from the compact core region at
15\,GHz become polarized while they propagate along the jet.  Beyond
2--3\,mas, the degree of polarization quickly rises to a level of
about 5\% as shown in the Figs.~\ref{fig:quantities_distance} and
\ref{fig:stack}.  Such a steep transition is unlikely to arise from a
smooth gradient in optical depth along the jet \citep[see, e.g.,
][]{Porth2011}.  Instead, the originally low degree of polarization
can be understood as a result of depolarization upstream of 2--3\,mas
in a foreground Faraday screen \citep[see, e.g.,][for the case of
3C~120]{Gomez2008}. In addition, beam depolarization due to multiple
polarized components close to the unresolved core might play an
important role given the complex and bent jet structure seen at
86\,GHz \citep{Schulz2012evn}.  Both possibilities are supported by
the strongly variable EVPAs in the inner 2\,mas with a dynamic range
of as large as 90\degr\ over 2\,mas (see
Fig.~\ref{fig:quantities_distance}, panel d). 

Here, we test for the effects of Faraday rotation as a
cause of those changes.\footnote{See Schulz et al, in prep., for a
  discussion of the influence of the complex bent jet structure within
  the inner 2\,mas on the EVPA changes reported here.}
Figure~\ref{fig:rm} shows the polarized flux-density distribution for
the same two quasi-simultaneous epochs at 15\,GHz (MOJAVE) and 43\,GHz
(BG data) for which the spectral index distribution was calculated in
Fig.~\ref{fig:spixmap}. In both maps, we are sensitive to polarized
intensity upstream of 2\,mas.

Panels c and d of Fig.~\ref{fig:rm} show a strongly changing and
double-peaked RM gradient along the jet ridge line (see the green open
squares in panel d). The RM changes rapidly between the core and
2\,mas distance with a maximum of $\approx$-8000\,rad/m$^2$ just
before 1\,mas. Following a dip down to 50--100\,rad/m$^2$, a second
peak appears around 1\,mas at $\approx$-5000\,rad/m$^2$. The RMs of
the two peaks correspond to EVPA rotations of roughly 180\degr\ and
40\degr, respectively, where the first can be compensated with the
ambiguity of $\pi$.  The high values obtained for the RM within this
region are consistent with recent results by \citet{Kravchenko2017}
for a large sample of jets.

The strong variations in the EVPAs observed upstream of 2\,mas
(Fig.~\ref{fig:quantities_distance}) also agree with the observed RM
inhomogeneities in that region. The observed inhomogeneous RM map (see
Fig.~\ref{fig:rm}) could be related with an external Faraday screen
\citep[e.g.,][]{Gomez2000,Gomez2008}, e.g., NLR clouds in the line of
sight \citep{ODea1989,Wardle1998}.

Nevertheless, the tentative transverse RM gradient
\citep[e.g.,][]{Asada2002,Asada2010,Croke2010,Hovatta2012,Gabuzda2014}
can also be an independent tracer for an underlying helical field.  The
detection of consistently negative circular polarization for the core
of 3C~111 in MOJAVE data \citep{Homan2006} gives an independent
argument for the presence of an ordered magnetic field configuration
at the core region.
\subsubsection{Intrinsic magnetic-field orientation}
Basic jet models  typically predict EVPAs, which are  either parallel 
or perpendicular to the overall jet flow.
Intrinsically parallel
EVPAs are allowed in cases with an axisymmetric helical magnetic
field depending on the viewing angle and the pitch of the helix
\citep{Lyutikov2005,Lyutikov2017}. \citet{Lyutikov2005} computed the expected EVPA configuration for jets
being threaded by a helical field of decreasing pitch angle towards
its fast spine. The larger spine emissivity causes observed EVPAs
to be always perpendicular to the jet. Similar conclusions are drawn from RMHD simulations for the selected jet velocities,
intrinsic pitch angles, and viewing angles, where a helical field is filled with plasma \citep[e.g.,][]{RocaSogorb2009,Gomez2016}.

Here, we investigate the implications of the measured RM for the
intrinsic EVPAs in the inner jet region. We assume for simplicity that
the RM shield is stable over time, we can tentatively infer the
intrinsic EVPAs (but see caveats discussed below). Figure~\ref{fig:rm}
shows the RM that is needed to have the intrinsic EVPAs of feature B
be aligned parallel to the jet and perpendicular to the jet,
respectively.  In the former scenario, we find a fairly good overall
agreement with the measured RM values (the region of the dip in RM is
not probed), while the scenario with perpendicular EVPAs predicts
systematically too high RM values upstream of $\sim$5\,mas from the core.
Our measurements thus show that underlying parallel EVPAs can explain
the observed RM gradient and the EVPA rotation over distance in the
inner-jet region upstream of about 1.5\,mas. The lack of a complete
coverage with RM data, however, does not allow us to conclude on the
bulk of the EVPA rotation at $>2$\,mas.

At this point, we have to note several caveats with respect to the
calculated RM values in Fig.~\ref{fig:rm}. First, the use of only two
frequencies introduces strong uncertainties on the measured RM values
and the simple $\lambdaup^2$-law may be broken close to the core
depending on the structure and geometry of the screen
\citep{Kravchenko2017}.  Furthermore, by measuring the RM coincident
for the well polarized feature B, we are sensitive only to the shocked
plasma of a single epoch and not the quiescent flow. We are therefore
lacking sufficient RM information for the entire parsec-scale jet both
in space and time.  For these reasons, we abstain from attempting to
directly correcting the measured EVPAs in
Fig.~\ref{fig:quantities_distance}.

\subsubsection{Component evolution and kinematics} The leading
components of both features are divided into three sub-components in
our modeling and show different behaviors: although B1 fades in
brightness very fast, A1 persists and keeps being bright across most
of the observing epochs. In both cases, the components A2 and B2
dominate in brightness further upstream. B2 becomes the leading
component of the feature B after B1 disappears. In addition, the
components A3 and B3 appear as bright features that fade rapidly and
disappear after four epochs, which follows a similar behavior to that
reported for component F in K08. The components E and F were studied
in terms of the hydrodynamical structure of the perturbation by
\citet{Perucho2008}.  These authors could successfully explain the
evolution of those two components, which were the only ones used to
model the region. Thus, although our current modeling shows a more
complex system of components, it seems to indicate that components A1
and B2 would correspond to E, whereas components A3 and B3 would
correspond to F in the interpretation given by
\citet{Perucho2008}. However, the richness of the structure revealed
in this work suggests that more detailed numerical simulations should
be performed, which is out of the scope of this paper.

\subsection{The region between 2--4\,mas}
Downstream of approximately 2\,mas from the core, the jet is
transversely resolved. Here, we observe a systematic smooth swing of
the overall EVPAs of the polarized features A and B as they propagate
with the jet flow. At the same time the jet overall structure is
remarkably straight and individual components show only small
deviations from their original trajectories (see Appendix A and the
component positions shown in the bottom panel of
Fig.~\ref{fig:stack}): this, in principle, leaves the interpretation
of the evolution in this region open to the possibility that some
components follow bent trajectories.  However, the amplitudes of these
changes and their contribution to the polarized emission are small so
that we neglect their influence on the large, observed EVPA rotation,
which has been claimed to be an explanation for other sources
\citep{Agudo2007,Molina2014,Lyutikov2017}. This also includes the
intrinsic near-perpendicular viewing angle (see Appendix~A), which
makes a projected geometrical EVPA rotation unlikely in the present
case. We can, however, not exclude effects due to an inhomogeneous
distribution of emitting plasma across the flow: Fig~\ref{fig:stack}
indicates that different areas of the jet cross-section are lighting
up at different times. These areas also include component trajectories
that do not follow the main stream. A discussion of related
higher-order effects, however, lies beyond the scope of this paper. In
this subsection we show that the results stated in this paper can be
explained by a recollimation shock within this region plus shearing of
the jet flow.

\subsubsection{A possible recollimation shock}
In Sect.~\ref{sec:tb} we showed evidence for unusual behavior at
around 3--4\,mas from the core, i.e., a sudden decrease of the feature
size $d$ at a distance of about 3\,mas, accompanied by an increase of
the brightness temperature $T_\mathrm{B}$ \citep[similar
to][]{RocaSogorb2009} and the polarized flux density of individual
model components. We recall that the power-law exponent describing the
evolution of the jet radius over distance changes from values smaller
to larger than 1 around 3\,mas. These observations can be interpreted
in terms of the presence of a recollimation shock at about this
distance, which was already suggested by K08.  Such recollimation
shocks are naturally forming extended structures in overpressured jets
as shown by numerical simulations
\citep[e.g.,][]{Gomez1997,Mizuno2015,Fromm2016,Marti2016}.

In the analysis that we have presented in Sect.~\ref{sec:tb}, we
provide estimates of the parameters $s$, $l$ and $\alpha$ involved in
the relation $s-l=n+b\,(1-\alpha)$. Since we can not constrain the
index $b$, we provide estimates for the resulting particle density
evolution in Sect.~\ref{sec:tb} for both a toroidal ($b=-1$) and an
axial ($b=-2$) field in a conically expanding jet
\citep{Pushkarev2017}. Independent of the choice for $b$, our results
show a steepening of the particle density described by a power-law
with the index $n$, possibly indicating the expected expansion after a
recollimation.

\subsubsection{The rotation of the EVPAs} 
The stacked polarization maps in Fig.~\ref{fig:stack} reveal a large
swing of about 180\degr\ already starting in the inner jet region,
related with the passage of the components. When assuming the presence
of a conical recollimation shock, the rotation could thus be explained
in terms of the bright features evolving into and out of the tip of
the conical shock. Unfortunately, RMHD simulations that tackle this
scenario are still missing.

Using calculations described in \citet{Cawthorne2006},
\citet{Agudo2012} could successfully explain the particular EVPA
distribution of a standing feature (associated to a recollimation
shock) in 3C~120 with a radial or Y-shaped pattern of EVPAs being
aligned along the central ridge and oblique EVPAs towards the
edges. \citet{Cawthorne2013} observe a similar pattern for the core of
S5~1803$+$784, equally arguing for the presence of a recollimation
shock from their modeling.  In steady situations, i.e., when there is
no interaction with a traveling component, these results demonstrate
the peculiar influence of conical shocks on the observed polarization
signatures. For 3C~111 the situation is clearly more complex, probably
involving a shock-shock interaction \citep[see, e.g.,][for a
relativistic hydrodynamics study]{Fromm2016}.

Contrary to the predictions by steady-state jets that form weak
recollimation shocks \citep{RocaSogorb2009,Gomez2016}, our RM data
favor intrinsically parallel EVPAs in the inner jet region
($\lesssim$1.5\,mas). Farther downstream the jet, \citet{Zavala2002}
also observed a gradient in RM, but at lower values smoothly ranging
from -800\,rad/m$^2$ to -200\,rad/m$^2$ between 3--5\,mas (see
Fig.~\ref{fig:rm})\footnote{These values correspond to a rotation of
  about 5--16\degr\ and describe a gradient of around 10\degr\
  rotation over 2\,mas distance in contrast to the much higher
  observed EVPA rotation rate of around 90\degr\ per 2\,mas in the
  region upstream of 2\,mas.}, which is consistent with typical values
of RM obtained for the MOJAVE sample \citep{Hovatta2012}.  Comparing
with our measurement at 2\,mas from the core, the RM appears to be
rather flat and stable along the outer jet. When extrapolating RM
found by \citet{Zavala2002} to 1.5\,mas and 6\,mas, Fig.~\ref{fig:rm}
suggests intrinsically perpendicular EVPAs in these regions.

The flatness and structural stability of the RM in the 2--4\,mas
region is in contradiction with it causing the large EVPA rotation.
We can tentatively address a simple explanation for this continuous
rotation of the EVPAs.

\begin{itemize}
\item Between $\sim$1.5 and 2\,mas, the RM departs from the one
  corresponding to an underlying parallel orientation of the EVPAs and
  approaches an intrinsically perpendicular orientation (see Fig.~13),
  because the measured RM translates into small rotations of the
  EVPAs. This initial rotation can be related to the shearing observed between
  components B2 and B4 between 1\,mas and 3\,mas (Fig. 10). The
  components are located at a central and peripheral position in the
  jet, respectively, and show different velocities, with the central
  component, B2, being faster. 86~GHz observations of the same flaring
  event leading to the B feature (Schulz et al., in prep.) show
  that these components are most probably individual ejections within
  the flare. These components are caught up by our observations,
  showing differential dynamics that possibly reveal a transverse
  structure of the jet velocity. The shearing of the jet plasma can
  explain the generation of a strong poloidal field component,
  resulting in a dominating perpendicular orientation of the
  EVPAs. This effect has been previously proposed by, e.g.,
  \citet{Laing1980,Laing1981}, \citet{Cawthorne1993b}, and
  \citet{Wardle1998}. The frequent observation of perpendicular EVPAs
  at the jet boundary and aligned EVPAs at the center seems to confirm
  a scenario where the sheared-layer stretching of the field lines
  dominates
  \citep{Dea1986,Attridge1999,Giroletti2004,Pushkarev2005}. Differential
  flows have also been observed in simulations of parsec-scale jets
  that show the stretching of lines at the jet shear layer
  \citep{RocaSogorb2009}, and in MHD simulations of kiloparsec-scale
  jets, which show this effect for the whole jet cross section
  \citep{Matthews1990,Gaibler2009,Huarte-Espinosa2011,Hardcastle2014}.

\item If strong planar shocks are traveling along the spine, we
  expect them to appear as bright components, such as B2, and show
  increased polarized emission with parallel EVPAs. At the interaction
  with the conical, standing shock, this regions becomes even brighter
  and can therefore explain the observed alignment of the EVPAs around
  3--4\,mas.

\item Finally, downstream of the recollimation shock, the flow
  progressively returns to the pre-shock situation, in which the
  polarized emissivity is dominated by the sheared region, with
  dominant perpendicular EVPAs.
\end{itemize}

Note that in the picture of an unsheared axisymmetric helical field
\citep[e.g.,][]{Lyutikov2005}, the observed intermediate EVPAs can not
be described solely by changes in the pitch angle of the helix (see
Appendix B) but potentially by the interaction of the propagating
shock with the recollimation shock -- a question that future RMHD
simulations need to answer. The dynamic nature and complexity of the
process must therefore be disassociated from the toy model shown in
Figure~\ref{fig:poldeg_helix}, which can successfully explain a
dichotomy between a parallel and perpendicular orientation but not the
continuous distribution of EVPAs.

\subsubsection{Estimate of the Mach number and magnetization}
If the change in the EVPA direction is caused by the presence of a
recollimation shock, we can obtain valuable insight into the jet
parameters by knowing the location of different standing shocks as
well as their transverse extent \citep{Marti2016}. There are
indications for standing features within the inner 0.5\,mas of the jet
(see Fig.~7).  Both Schulz et~al. (in prep.) and \citet{Jorstad2017}
find similar standing features at 86\,GHz and 43\,GHz,
respectively. The evidence collected indicates the presence of a first
recollimation shock in close vicinity to the millimeter core and a
second one at around 3\,mas, that is, the detection of multiple
recollimation shocks in 3C~111, similar to, e.g., BL~Lac
\citep{Gomez2016,Mizuno2015}, CTA~102 \citep{Fromm2013,Fromm2016} or
3C~120 \citep{LeonTavares2010,RocaSogorb2010,Agudo2012}. Based on
these evidence, we can estimate the magnetosonic Mach number and put
qualitative constraints on the magnetization of the flow \citep[see
also][for independent estimates of the jet
magnetization]{Nokhrina2015}.

\begin{figure*}
  \centering
  \includegraphics[width=\textwidth]{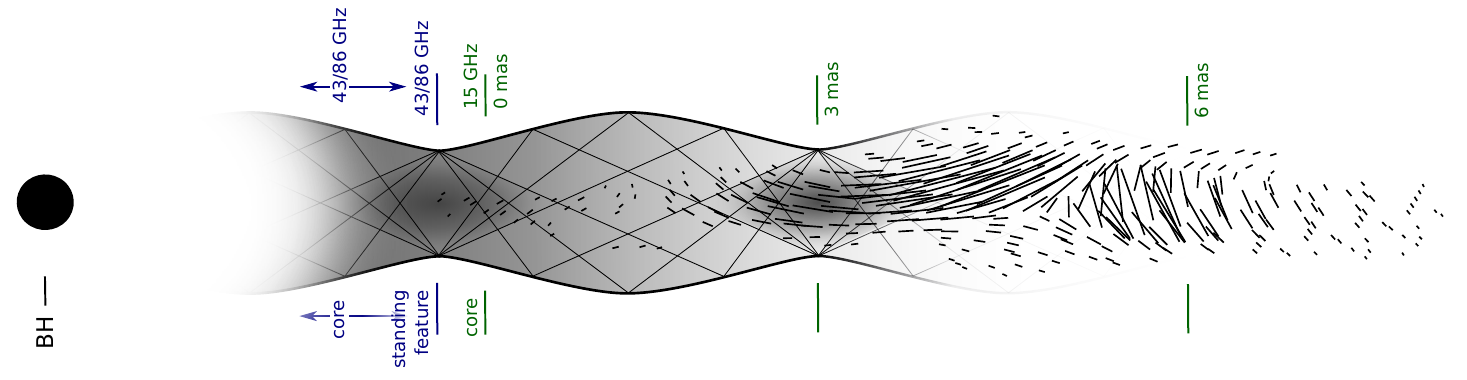}
  \caption{Illustration of the relevant observed features mixed with
    the theoretical prediction of a continuously expanding and
    recollimating flow \citep[e.g.,][]{Daly1988,Gomez1997} with an
    entrained particle distribution of downstream decreasing
    density. The straight lines indicate Mach cones for different
    speeds of the flow. Standing features observed at 43\,GHz and
    86\,GHz are interpreted as the first recollimation shock
    downstream of and in close vicinity to the cores at these
    frequencies. Its distance to the black hole is unknown for
    3C~111. At 15\,GHz, we observe a second recollimation shock at a
    distance of 3\,mas from the core component at that frequency. We
    detect no signs for further recollimations. Overlaid, we plot the
    EVPAs from the stacked image in Fig.~\ref{fig:stack}. The
    orientation of these EVPAs appears to be perpendicular to the jet
    in absence of the recollimation and parallel on top of it.}
  \label{fig:recoll_sketch}
\end{figure*}

We can use the inferred distance of $\sim$3\,mas between the 15\,GHz
core and the downstream recollimation as an approximation to the
distance between two shocks (see
Fig.~\ref{fig:recoll_sketch})\footnote{The shift between the cores at
  43\,GHz/86\,GHz and 15\,GHz due to synchrotron self-absorption is
  negligible (see Fig.~12).}.  The deprojected separation corresponds
to 7.6--18.6\,pc, depending on the jet viewing angle (see Appendix
A). This translates to $2.2$--$5.3 \times
10^{5}\,r_\mathrm{s}$. Although this recollimation shock is likely not
the first one along the jet, the order of magnitude compares well with
distances of recollimation shocks of around $10^{5}\,r_\mathrm{s}$ for
the examples of M87, CTA~102 or BL~Lac.

\citet{Marti2016} have performed RMHD simulations to study the
internal structure of overpressured jets that form a series of
recollimation shocks, covering a wide range of the jet magnetization
and internal energy. They report on the correlation between the
magnetosonic mach number $\mathcal{M}_\mathrm{ms}$ and the
half-opening angle of the flow $\tan{\phi}=2R/D \sim
1/\mathcal{M}_\mathrm{ms}$ (with $R$ the jet radius at its maximum
expansion and $D$ the distance between two recollimation shocks). We
find that the jet of 3C~111 is resolved at 3\,mas (see Fig.~11) with a
maximum FWHM of approximately $\sim$1\,mas. This value gives a lower limit
to the jet expansion radius because 1) the jet flow can be wider than
the visible radio jet, and 2) the location of the maximum jet
expansion is unknown. Therefore, from this value and the distance
between shocks, we obtain an upper limit of
$\mathcal{M}_\mathrm{ms}$. The values that we obtain are
$\mathcal{M}_\mathrm{ms}\lesssim 7.6$--$18.6$.

The magnetosonic Mach number itself is defined as
\begin{equation}
  \label{eq:machnumber}
  \mathcal{M}_\mathrm{ms}=\frac{\Gamma_\mathrm{j}\,v_\mathrm{j}}{\Gamma_\mathrm{ms}\,c_\mathrm{ms}}
\end{equation}
with $\Gamma_\mathrm{j}$, the Lorentz factor of the jet, and
$\Gamma_\mathrm{ms}$, the Lorentz factor associated with the
magnetosonic speed $c_\mathrm{ms}$, and $v_\mathrm{j}$, the jet
speed. The observed superluminal components and the large internal jet
speed, estimated in Appendix~A, allows us to approximate
$v_\mathrm{j}\simeq c$. Thus, using $c=1$,
$\mathcal{M}_\mathrm{ms}=\Gamma_\mathrm{j}/(\Gamma_\mathrm{ms}\,c_\mathrm{ms})$. The
values obtained for the magnetosonic Mach number ($\lesssim$7.6--18.6)
allow us to approximate $\Gamma_\mathrm{ms}\simeq 1$ as otherwise the
bulk Lorentz factor needs to be large ($\leq 20$).  Finally, we obtain
$\mathcal{M}_\mathrm{ms}=\Gamma_\mathrm{j}/\,c_\mathrm{ms}$.  Under
these assumptions and considering Fig.~15 from \citet{Marti2016}, the
jet will be kinetically dominated for bulk Lorentz factors
$\Gamma_\mathrm{j}\gtrsim 4.8 - 11.6$ (depending on the deprojected
distance between the shocks, see above). These values are of the order
of those observed (see Appendix A). When we insert the lower limit on
the jet speed $v_\mathrm{min}\sim 0.976\,c$ close to the inclination
angle of $\theta_\mathrm{min}=10$\degr\ and the corresponding upper
limit on the Mach number $\mathcal{M}_\mathrm{ms, max}=18.6$ into
Eq.~\ref{eq:machnumber}, we can establish a lower limit of
$c_\mathrm{ms, min}=0.23$. Independent measurements of the jet speed
and inclination angle by \citet{Jorstad2005} and \citet{Hovatta2009}
result in magnetosonic speeds of $c_\mathrm{ms}=0.38$ and
$c_\mathrm{ms}=0.53$, respectively, favoring a jet being kinetically
dominated or just at the transition of being Poynting-flux dominated
\citep{Marti2016}.

\subsection{The jet beyond 4\,mas}
The region downstream of the recollimation is characterized by a
smooth and consistent 90\degr\ eastward swing of the EVPAs of both
features A and B from being aligned with the jet at the recollimation
to a perpendicular orientation. This swing has been explained in the
previous section with a helical field stretched out towards being
poloidally dominated due to a velocity shear caused by the bulk flow
of the jet.  Such a field would give rise to EVPAs
being predominantly oriented perpendicular to the jet, which is also
predicted by numerical simulations
\citep[e.g.,][]{Huarte-Espinosa2011,Hardcastle2014}.  Sample studies,
moreover, have revealed transverse EVPAs to be typical features for
FR~II jets \citep{Bridle1984a,Hardcastle1997,Gilbert2004}. Also,
results by \citet{Lister2000} and \citet{Kharb2008} suggest that
overall transverse EVPAs are found in presumable quiescent quasar
jets, as opposed to shocked quasars, where a larger fraction of
parallel EVPAs has been found.  Furthermore,
\citet{Perucho2005,Marti2016} have shown that the presence of a shear
layer slows down the growth of Kelvin-Helmholtz or current-driven
instabilities, providing the jet with increased stability, as expected
for FRII jets like 3C~111. Observational evidence is provided, e.g.,
by \citet{Kharb2008,Gabuzda2014}, and, in the case of 3C~111, by
K08. This effect reinforces the dominance of transverse EVPAs (before
and) after the interaction of the traveling features with the standing
shock.

Beyond the suggested recollimation shock, the expansion rate increases to
values larger than one (see Section 3.2). Also, the density gradient becomes steeper,
as obtained from the evolution of the component diameter and
brightness temperature with distance. These results seem to indicate a
freely expanding jet, possibly because of a steepening of the ambient
pressure profile, compatible with the interpretation
of foregoing MOJAVE data by K08. We also find a number of new
components that form in the wake of the leading components A\,2 and
B\,2. These new components can be interpreted as trailing features
\citep{Agudo2001,Jorstad2005}. The oscillation that is produced in the
rear of the leading component can affect the entire jet cross-section
and can be strong enough to produce conical shocks, which themselves
propagate downstream and are sufficiently bright to be detected
\citep{Agudo2001,Mimica2009}. 

\section{Summary and Conclusions}
\label{sec:summary}
We have investigated the complex evolution of the 3C~111 jet flow that
originated from millimeter-wavelength outbursts just before 2006
(outburst A), 2008 (outburst B), and 2009 (outburst C). As part of the
flow, individual apparent superluminal features were tracked and
characteristic parameters were recorded both in time and core
distance, namely, the (polarized) flux density, brightness temperature
and feature size. This dedicated study was facilitated by the
increased density of the MOJAVE monitoring of 3C~111 and the
availability of polarization information for all epochs.  We summarize
our key observational results as follows:

\begin{itemize}
\item \textit{Upstream of $\sim$2\,mas} the EVPAs are strongly
  variable with time. This variability has a potential de-polarizing
  effect, which can naturally be explained by the observed strong and
  inhomogeneous RM distribution across that region. Also, beam
  depolarization plays a role in that regard. When accounting for possible
  rotation, we infer intrinsically parallel EVPAs upstream of
  $\sim$1.5\,mas. 

  \textit{Within $\sim$1.5--2\,mas}, the Faraday-corrected EVPAs
  become transverse to the jet. We interpret this as produced by a
  shear at the boundary layers of the jet flow, i.e., a differential
  transverse velocity structure leading to a dominant axial magnetic
  field in this region.  Such a shear is supported by the peculiar
  kinematics of the jet components B\,2 and B\,4, which appear as the
  continuation of two components observed at 86\,GHz for this same
  flaring event (Schulz et al., in prep.).

\item \textit{At 2--4\,mas} we find indications for a recollimation
  shock based on a sudden increase of the brightness temperature
  accompanied with a decrease in feature size for individual
  components. Also, the polarized intensity reaches large values in
  this region. Beyond this region, the power-law index describing the
  growth of the components increases moderately, while the particle
  density gradient steepens significantly. Our estimates of the
  magnetosonic Mach number based on the distance between standing
  components \citep{Marti2016} put the parsec-scale jet well into the
  kinetically dominated regime, or just at the transition to being
  Poynting-flux dominated. Based on archival information, we conclude
  on a low and relatively flat RM gradient downstream of
  $\sim$2\,mas. We can therefore not explain the bulk of the EVPA
  rotation with a changing Faraday screen. Instead, our observation of
  aligned EVPAs at the recollimation as well as intermediate angles in
  between can likely be the result of the interaction of shocked
  ejecta with a standing shock that may be able to further enhance the
  toroidal field component.
  
\item \textit{Beyond 4\,mas} the stacked polarized maps indicate an
  extended region with significant levels of polarization (associated
  to the passing components) with an overall trend of EVPAs turning
  back into a transverse orientation. Similar to the situation around
  2\,mas, we propose a boundary layer interaction for causing a
  dominant axial field component following the possible reacceleration
  of the jet flow at the spine with respect to that at the
  boundaries. 3C~111 therefore fits well into the common frame of
  (quiescent and unperturbed) FR~II jets, where jets typically show a
  dominance of transverse EVPAs.  Furthermore, we also detect trailing
  features in this region that may represent secondary pinch mode
  perturbations in the wake of the leading shock.

\end{itemize}

3C~111 turns out to be a uniquely well suited source to probe the
physics of the innermost parsec-scale jet by carefully mapping the
distribution and long-term evolution of (polarized) intensity over
tens of parsec downstream distance. We find that the peculiar behavior
and large rotation of the EVPAs ($\sim$180\degr\ across a distance of
4\,mas, i.e., $\lesssim$25\,pc deprojected distance) may be related to
a deviation from the quiescent situation of a sheared jet flow, i.e.,
the passage of shocks through a recollimation shock. Such an
interaction, however, lacks corresponding RMHD simulations with full
radiative output including polarization. We therefore propose the
presented results as observational reference for future simulations.

\begin{acknowledgements}
  TB thanks for fruitful discussions with M. Lyutikov, E. Kravchenko
  and D.~Gabuzda for her input that helped to improve the paper. We
  thank the National Radio Astronomy Observatory, a facility of the
  National Science Foundation operated under cooperative agreement by
  Associated Universities, Inc.; this research has made use of data
  from the MOJAVE database that is maintained by the MOJAVE team
  (Lister et al., 2009) and IRAM, which is supported by INSU/CNRS
  (France), MPG (Germany) and IGN (Spain). This work has benefited
  from research funding from the European Community's Seventh
  Framework Programme. The Submillimeter Array is a joint project
  between the Smithsonian Astrophysical Observatory and the Academia
  Sinica Institute of Astronomy and Astrophysics and is funded by the
  Smithsonian Institution and the Academia Sinica. We also thank
  J. E. Davis for the development of the \texttt{slxfig} module that
  has been used to prepare the figures in this work.  We made use of
  ISIS functions provided by ECAP/Remeis observatory and MIT
  (\url{http://www.sternwarte.uni-erlangen.de/isis/}) as well as the
  NASA/IPAC Extragalactic Database (NED), which is operated by the Jet
  Propulsion Laboratory, California Institute of Technology, under
  contract with the National Aeronautics and Space
  Administration. M.P. acknowledges financial support from the Spanish
  Ministerio de Economía, Industria y Competitividad (grants
  AYA2013-40979-P, and AYA2013-48226- C3-2-P) and from the local
  Valencian Government (Generalitat Valenciana, grant
  Prometeo-II/2014/069). I.A. acknowledges support by a Ram\'on y
  Cajal grant of the Ministerio de Econom\'ia y Competitividad
  (MINECO) of Spain. A.B.P.\ and Y.Y.K.\ acknowledge partial support
  by the Russian Foundation for Basic Research (grant 17-02-00197), by
  the government of the Russian Federation (agreement 05.Y09.21.0018),
  and by the Alexander von Humboldt Foundation.  E.R. acknowledges
  partial support by the Spanish MINECO project AYA2012-38491-C02-01
  and by the Generalitat Valenciana project PROMETEOII/2014/057. T.S.
  was funded by the Academy of Finland projects 274477 and 284495. The
  research at the IAA-CSIC was supported in part by the MINECO through
  grants AYA2016-80889-P, AYA2013-40825-P, and AYA2010-14844, and by
  the regional government of Andaluc\'{i}a through grant P09-FQM-4784.
\end{acknowledgements}

\newcommand{\noop}[1]{}

\appendix

\section{Constraints on the viewing angle and the intrinsic speed}
\label{sec:inclination}
\begin{figure}
    \centering
    \resizebox{\hsize}{!}{\includegraphics{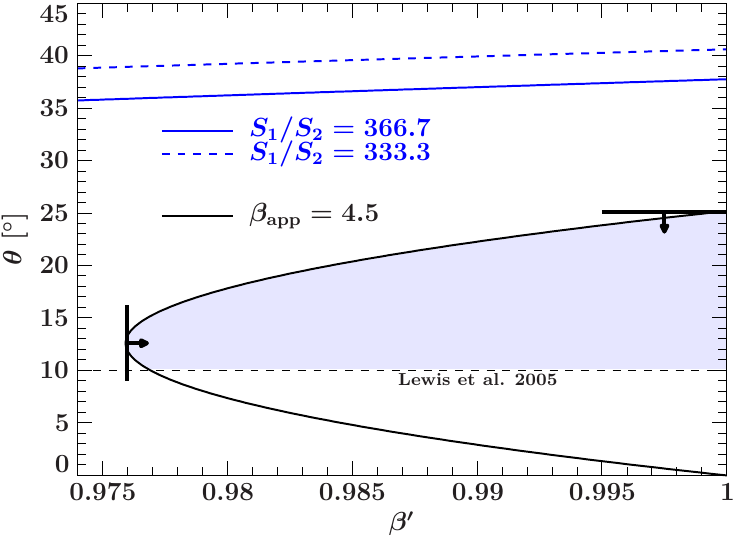}}
    \caption{Constraints on the intrinsic jet velocity $\beta^\prime$
      and the inclination angle $\theta$ based on the average measured
      apparent speed $\beta_\mathrm{app}=4.5$ for the leading
      components B\,1--B\,4 of feature B. Two estimates for the
      jet-to-counterjet flux-density ratio $S_{1}/S_{2}$ define
      further constraints drawn as blue lines. The blue-shaded region
      highlights the allowed parameter space, the thick lines with
      corresponding arrows define the lower and upper limits on the jet
      speed and the inclination angle.}
    \label{fig:beta_theta}
  \end{figure}

We estimate the viewing angle $\theta$ and the intrinsic jet speed
$\beta^\prime$ using the relation for the apparent speed
  \begin{equation}
    \beta_\mathrm{app}=\frac{\beta^\prime \sin{\theta}}{1-\beta^\prime \cos{\theta}}.
    \label{eq:beta_app}
  \end{equation}
  We choose the observables measured for the components B\,1--B\,4
  that describe the jet just downstream of the core in 2008. Their
  average apparent proper motion is 1.27\,mas/yr, translating to
  $\beta_\mathrm{app}=4.5$ with $1\,\mathrm{mas}=1.08\,\mathrm{pc}$
  when applying the latest cosmological parameters from
  \citet{Planck2016}.  We insert $\beta_\mathrm{app}$ in
  Eq.~\ref{eq:beta_app} and plot the solution for
  $\theta\,(\beta^\prime)$ in Fig.~\ref{fig:beta_theta} as solid,
  black line. This line both defines the lower limit on the jet speed
  $\Gamma^\prime \geq (1+\beta_\mathrm{app})^{1/2}$, i.e.,
  $\beta^\prime \geq 0.976$ and the upper limit on the viewing angle
  $\theta \leq 2\arctan{\beta_\mathrm{app}^{-1}} =25.1$\degr.
  Independent estimates of the viewing angle were derived by also
  measuring the Doppler factor via the decline time of ejected knots
  in addition to their apparent speed \citep{Jorstad2005}. They
  equally state a possible range of $\theta\sim 10$--$25$\degr\ for a
  number of observed knots with a weighted average of 18.1\degr. This
  range is consistent with an upper limit of 20\degr\ found by
  \citet{Oh2015}. We adopt the lower limit of 10\degr\ estimated by
  \citet{Lewis2005} for the large-scale morphology, although lower
  values cannot be excluded based on parsec-scale
  kinematics. Extending on the method used by \citet{Jorstad2005},
  \citet{Hovatta2009} estimate the Doppler factor based on
  $T_\mathrm{b}$ variability and find an inclination of
  about 15.5\degr\ when considering an apparent component speed of
  $\beta_\mathrm{app}\sim 5.9$.  Both proposed viewing angles are well
  consistent with our estimate.

  Figure~\ref{fig:beta_theta} also shows upper limits based on two
  estimates of the jet-to-counterjet ratio
\begin{equation}
    \frac{S_1}{S_2}=\left(\frac{1+\beta^\prime \cos{\theta}}{1-\beta^\prime \cos{\theta}}\right)^{2-\alpha},
    \label{flux_ratio}
  \end{equation}
  where $\alpha$ is the spectral index, which is calculated for the
  components B\,1--B\,4 between two MOJAVE epochs at 2008-09-12 and
  2009-01-30 as well as quasi-simultaneous archival observations at
  43\,GHz. We find values of $\alpha=-0.75$ and $\alpha=-0.92$,
  respectively. The integrated flux density of these components at
  both epochs are $S_{1}=2.2$\,Jy and $S_{1}=1.4$\,Jy with
  corresponding estimates on the counter-jet flux density of
  $S_{2}=6.0$\,mJy and $S_{2}=4.2$\,mJy, respectively, when assuming
  an unresolved counter-jet, if it was detected. Both values of
  $S_{2}$ are upper limits and correspond to $10\sigma$ of the
  background rms, which is a rather conservative choice.  Our
  estimates using the flux ratios exclude viewing angles above
  $\sim$35\degr, while better constraints on the counter-jet flux
  density could help to further constrain the maximum viewing angle of
  about 25\degr\ set by the measured apparent speed.

  \begin{figure}
    \centering
    \resizebox{\hsize}{!}{\includegraphics{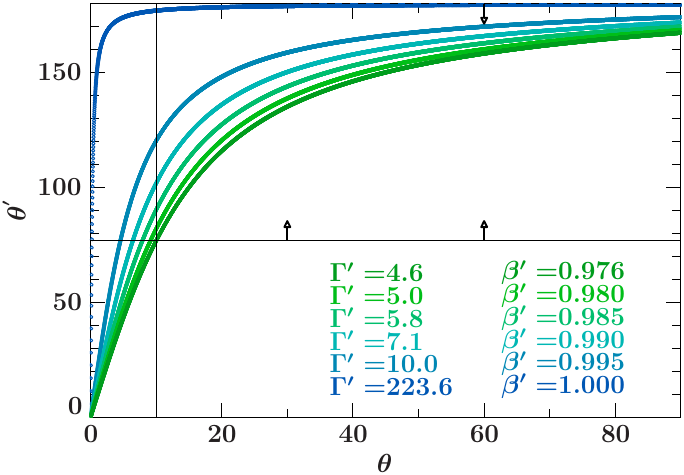}}
    \caption{Constraints on the intrinsic viewing angle
      $\theta^\prime$ based on previous estimates of the viewing angle
      in the observers frame and the intrinsic speed
      $\beta^\prime$. We plot the relation $\theta^\prime (\theta)$
      for a range of allowed values of $\beta^\prime$.  The lower
      limit of $\theta =10$\degr\ at a speed of $\beta^\prime\sim
      0.976$ sets the lower limit on $\theta^\prime$.}
    \label{fig:rf_angle}
  \end{figure}
  Using the estimate on $\theta$, we can also constrain the intrinsic
  viewing angle to $\theta^\prime$ by
  using the Lorentz-transformation
  \begin{equation}
    \label{eq:rf_angle}
    \cos{\theta^\prime} = \frac{\cos{\theta}-\beta}{1-\beta \cos{\theta}}.
\end{equation}
In Fig.~\ref{fig:rf_angle} we plot this function for a number of
values of $\beta^\prime\gtrsim 0.976$. The lower limit on the viewing
angle of 10\degr\ at a speed of $\beta^\prime\sim 0.976$ establishes
the lower limit of $\theta^\prime \gtrsim 76.8$\degr. When instead
inserting the estimates by \citet{Jorstad2005} and
\citet{Hovatta2009}, who additionally constrain the Doppler factor
from flux and $T_\mathrm{B}$-variability arguments, we find
$\theta^\prime\sim 108$\degr\ for $\Gamma=4.4$, $\theta=18.1$\degr\
\citep{Jorstad2005}, and $\theta^\prime\sim 129$\degr\ for $\Gamma=7.7$
and $\theta=15.5$\degr\ \citep{Hovatta2009}.

\section{Polarization signatures and a possible helical magnetic
  field}
\label{sec:toymodel}
The lack of numerical simulations that tackle the interactions between
moving and stationary shocks in a RMHD setup, forces us to discuss our
results using simplifying assumptions. \citet{Lyutikov2005} consider
an axisymmetric magnetic field constraining a hollow cylindrical jet and
infer corresponding polarization properties.

\begin{figure}
  \centering
  \resizebox{\hsize}{!}{\includegraphics{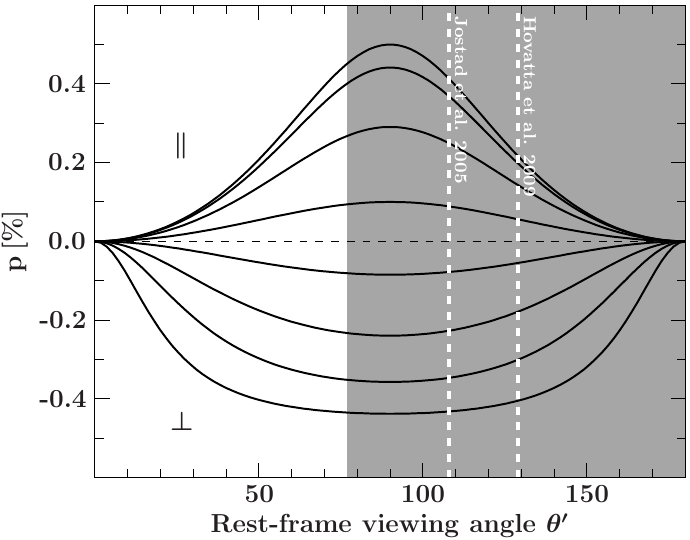}}
  \caption{Degree of polarization for a range of rest-frame viewing
    angles $\theta^\prime$ onto axial-symmetric helical magnetic
    fields in a hollow cylindrical geometry. The estimated range of
    $\theta^\prime$ for 3C~111 is highlighted as gray-shaded region
    with two independent estimates as white, dashed lines based on
    inclination angles stated in the literature. Negative values of
    $p$ correspond to an integrated EVPA perpendicular to the jet
    axis, positive values result in parallel EVPAs. The figure shows a
    range of intrinsic pitch angles $\psi^\prime$ with steps of
    10\degr\ between 20\degr--90\degr\ from the bottom to the top.  This
    example corresponds to the analytical expression provided for a
    particle distribution $N(E)\,dE\sim E^{p}\,dE$ with $p=3$ in Eq.~21 by
    \citet{Lyutikov2005}.}
  \label{fig:poldeg_helix}
\end{figure}
Figure~\ref{fig:poldeg_helix} shows the expected, integrated degree of
polarization for such an axisymmetric helical field in a hollow-jet
geometry.  We plot multiple solutions for a range of intrinsic pitch
angles $\psi^\prime$ between $20\mathrm{\degr}$--$90\mathrm{\degr}$
(from the bottom to the top).  In the expression given by
\citet{Lyutikov2005} negative and positive values correspond to
orthogonal and aligned EVPAs, respectively.  Above
$\psi^\prime=50$\degr, the field is getting toroidally dominated
featuring EVPAs aligned with the jet. The degree of polarization will
be maximal for a pure toroidal field with $\psi^\prime=90$\degr\ a
pure axial field for very low pitch angles. In summary, the EVPA
distribution would appear as bimodal, depending on the dominant
component, i.e., the pitch angle of the field.

A 90\degr\ swing in the observed EVPA can thus occur by changes in the
intrinsic pitch angle. We warn the reader that this result is obtained
using a determinate electron distribution exponent that allows for an
analytical solution of the equations. Changes in the EVPA orientation
can also occur with changing viewing angle only for a particle
distribution of lower index ($p<3$) and for a small range of
pitch angles \citep{Lyutikov2017}.  In addition, studies of large
samples of blazars seem to confirm this quasi-bimodal distributions
of EVPAs \citep{Bridle1984a} with the majority of VLBI knots in BL~Lac
objects showing aligned EVPAs
\citep{Gabuzda1994,Gabuzda2000,Lister2000}.

In Sect.~\ref{sec:inclination}, we determine a lower limit on the
intrinsic viewing angle of $\theta^\prime = 76.8$\degr\ based on the
range of viewing angles in the observers frame of
$\theta=10$\degr--$25.1$\degr. The allowed range for $\theta^\prime$
with respect to our kinematics is shown as gray-shaded region in
Fig.~\ref{fig:poldeg_helix}. The independent estimates of
$\theta^\prime \sim 108$\degr\ and $\theta^\prime \sim
129$\degr\ based on inclination angles proposed by \citet{Jorstad2005}
and \citet{Hovatta2012} are marked as white dashed lines. Our
observations of a polarization degree of around 10\%--20\% would be
consistent with this model for pitch angles around 60--70\degr\ in
case of a toroidal field or for pitch angles around 40--50\degr\ in
case of an axial field.

\section{Evolution of the model components in \textit{x/y}-space}
\begin{figure}
    \centering
    \resizebox{\hsize}{!}{\includegraphics{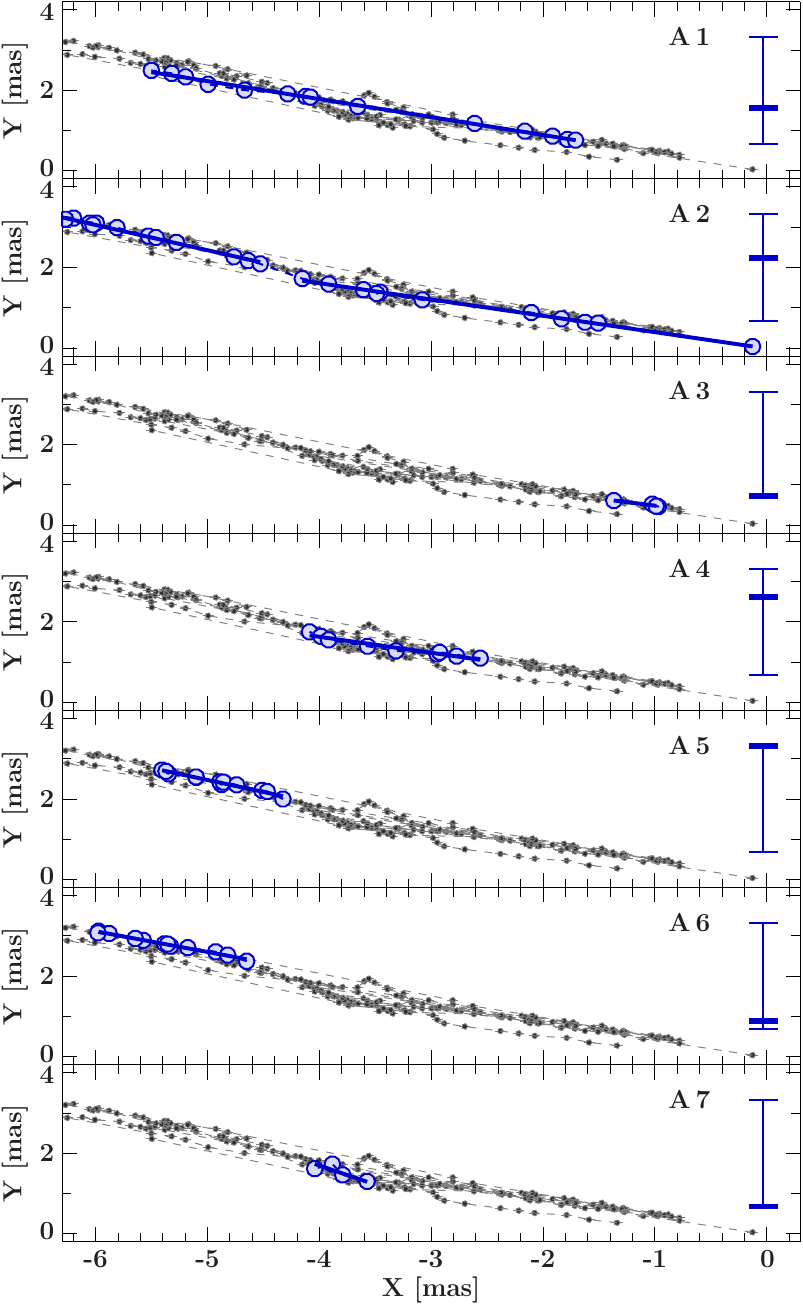}}
    \caption{Model component positions in \textit{x/y}-space for the
      feature A. In each panel, all model components are shown as
      gray circles in the background. The colorized bar on the right of
      each panel denotes the average polarized flux of each component
      within the minimum/maximum fluxes of all components of the
      feature A, i.e., 2.2/11.2\,mJy.}
    \label{fig:xy_A_appendix}
  \end{figure}

\begin{figure}
    \centering
    \resizebox{\hsize}{!}{\includegraphics{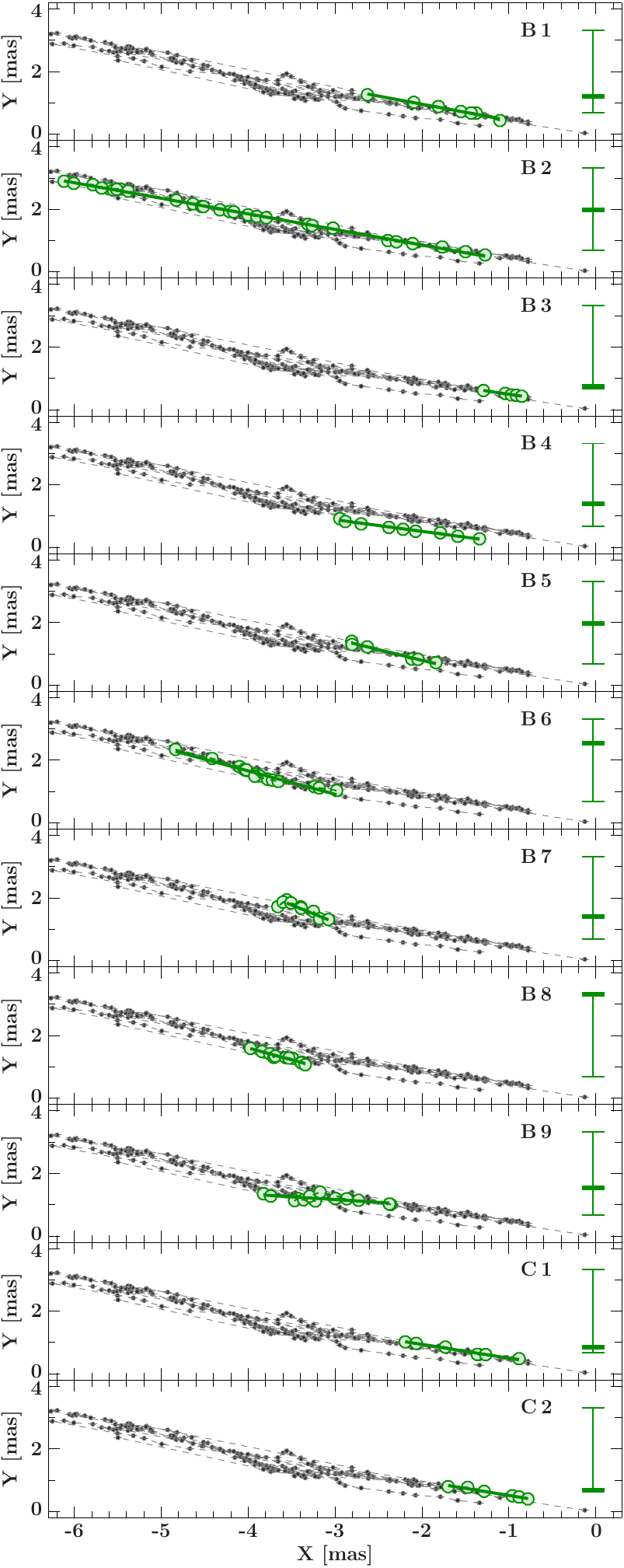}}
    \caption{Model component positions in \textit{x/y}-space for the
      feature B. In each panel, all model components are shown as
      gray circles in the background. The colorized bar on the right of
      each panel denotes the average polarized flux of each component
      within the minimum/maximum fluxes of all components of the
      feature B, i.e., 4.6/17.8\,mJy.}
    \label{fig:xy_B_appendix}
  \end{figure}

\end{document}